\documentclass[draft]{agujournal2019}
\usepackage{url} 
\usepackage{soul}
\usepackage{amsmath}
\usepackage{esint}

\draftfalse

\journalname{Journal of Advances in Modeling Earth Systems}

\begin{document}

%
%


\title{How Convective Mass Flux Responds to Environmental Humidity}

%
%




\authors{Seth D. Seidel\affil{1}, Nathan P. Arnold\affil{1}, and Brandon Wolding\affil{2,3}}


\affiliation{1}{Global Modeling and Assimilation Office, NASA/GSFC}
\affiliation{2}{Cooperative Institute for Research in Environmental Sciences, University of Colorado Boulder}
\affiliation{3}{NOAA/Earth System Research Laboratory}




\correspondingauthor{Seth D. Seidel}{seth.seidel@nasa.gov}



 \begin{keypoints}
 \item	Convective mass flux transitions to deep inflow and increases quasi-exponentially as lower tropospheric relative humidity increases.
 \item	Greater lower tropospheric humidity is not associated with greater population-mean buoyancy in convective updrafts.
 \item	These phenomena are the result of increasing survival and dilution among convectively active parcels of air.
 \end{keypoints}

%
%

%
%


\begin{abstract}
Our goal in this study is to characterize the relationship between lower tropospheric environmental humidity and convective mass flux in the tropics. To do so, we have created gridded convective mass flux datasets from five global storm-resolving models (GSRMs). We have three principal findings. First, in humid environments, mass flux increases with height from the surface through the depth of the lower free troposphere, forming a ``deep-inflow". In dry environments, mass flux does not increase with height in the lower free troposphere. Second, mid-tropospheric mass flux increases nonlinearly with increasing lower tropospheric humidity, resembling a widely reported pickup in tropical precipitation. Third, increased lower tropospheric humidity is associated with reduced deep convective updraft buoyancy. To interpret these findings, we employ a simple three-equation parcel model with stochastic entrainment. The parcel model suggests that the response of convective mass flux to lower tropospheric humidity is governed by two effects: (1) survival, in which a greater share of entraining parcels ascend rather than detrain with greater humidity; and (2) dilution, in which the average entrainment rate among surviving parcels increases with environmental humidity. Together, survival and dilution account for the three mass flux responses to humidity.
\end{abstract}

\section*{Plain Language Summary}
This study aims to quantify and understand the rate at which air mass ascends within cumulus and cumulonimbus clouds in the tropics. This rate is known as \textit{convective mass flux}. Using fine-scale supercomputer simulations of Earth's atmosphere, we find that convective mass flux is extremely sensitive to humidity in the lowest few kilometers of the atmosphere. Greater humidity leads to greater convective mass flux. We then test whether environmental humidity increases mass flux by making cloudy air more buoyant (i.e., less dense relative to its surroundings). We reject this hypothesis, finding instead that the opposite occurs: Greater humidity is associated with cloudy air which is \textit{less} buoyant. To make sense of these results, we use a simple set of equations to simulate cloudy air as it rises and ingests dry environmental air. As the environment becomes more humid, cloudy air may absorb a greater mass of environmental air without drying out and halting its ascent. This causes mass flux to increase with humidity and causes convective clouds' average density to become more like that of their environment.

%
%

%


%
%
%
%

\section{Introduction}

The tropical atmosphere is divided between broad regions of subsidence and narrow regions of updrafts within convective clouds. This cumulus convection constitutes the vast majority of upward mass transport and plays several mportant roles in tropical dynamics: The upward branches of the Hadley and Walker circulations are composed of convective updrafts. Most precipitation in the tropics is the result of convective updrafts. The upward transports of water and energy are directly related to the magnitude of this mass transport. And extensive cirrus clouds appear in the upper troposphere, where the mass transport declines with height such that air (and cloud water) must be expelled from convective cores \cite{Hartmann2002,Bony2016,Seidel2022}. We typically define the quantity \textit{convective mass flux} ($M$) as the upward transport of air mass within convective clouds per unit area. In mathematical form:

\begin{linenomath*}
\begin{equation}
M = \sigma_{u} * \rho * w_{u}
\end{equation}
\end{linenomath*}

where $\sigma_{u}$ is the area fraction of updrafts within some larger domain, $\rho$ is the density of air, and $w_{u}$ is the average vertical velocity in updrafts.

Convective mass flux is difficult to study at a global scale. This difficulty is partly due to a paucity of observations, for in-cloud velocity is often difficult to observe except when using ground-based radar installations, which have limited coverage in the tropics \cite{Giangrande2016,Savazzi2021}. Existing satellite-based estimates of mass flux are subject to large uncertainties and are built upon several physical assumptions regarding convective clouds \cite{Jeyaratnam2021}. On the other hand, general circulation models (GCMs) are typically run at such coarse horizontal resolutions that cumulus parameterizations are required to represent the effects of convection. However, the steady decline of computing costs have recently allowed modeling centers to deploy global storm-resolving models (GRSMs). These models have  a typical grid spacing of 2-5 km, so that deep convection may be permitted at the resolved scale rather than parameterized. We shall take advantage of this new tool. We will quantify convective mass flux in the GSRMs and examine its response to environmental conditions. In doing so, we hope not only to advance our understanding of convection in the real atmosphere, but also to provide an informative comparison across GSRMs.

\begin{figure}
    \centering
    \includegraphics[width=1\linewidth]{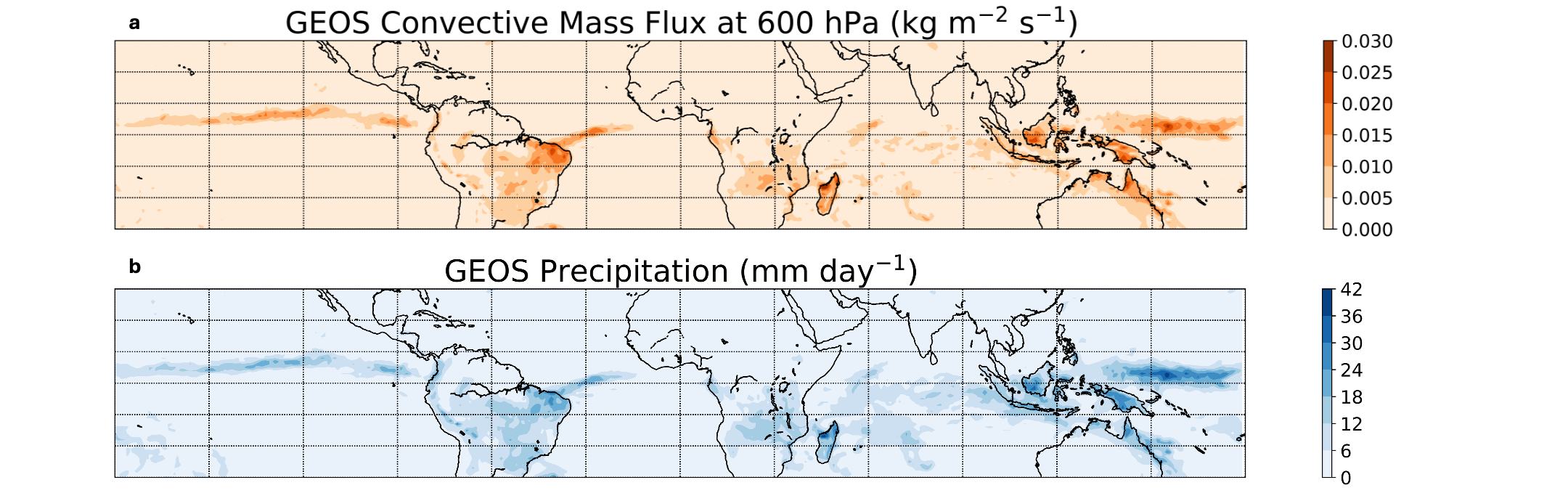}
    \caption{(a) Convective mass flux simulated by the NASA GEOS GSRM for February 2020. (b) Precipitation simulated by the NASA GEOS GSRM for February 2020.}
    \label{fig:maps}
\end{figure}

To select a framework for studying convective mass flux, we look to past studies of tropical precipitation. Figure \ref{fig:maps} shows precipitation and mid-troposphere convective mass flux in the tropics, as simulated by NASA's GEOS GSRM. The precipitation and mass flux fields are strikingly similar at a glance. This is not surprising, as heavy precipitation in the tropics is nearly always the result of convective updrafts. Given that precipitation and mass flux are empirically and theoretically linked, it makes sense to study global-scale convective mass flux using a framework that has proved useful for precipitation. Our framework of choice is the precipitation response to column humidity. As humidity increases in the tropical troposphere, mean precipitation increases sharply, indicating the onset of strong convection. This ``precipitation pickup" has been found in observations \cite{Bretherton2004,Neelin2009,Ahmed2015,Wolding2020} as well as in models \cite{Igel2017,Igel2017a,Rushley2018}. Our first goal in this study is simply to establish \textit{what} is the convective mass flux response to humidity in the GSRMs.

Our second goal is to understand \textit{why} convective mass flux responds to humidity in the way that it does. The precipitation pickup phenomenon has inspired a body of theory relating the precipitation rate to the buoyancy of an entraining plume \cite{Holloway2009,Ahmed2018,Adames2021}, which necessarily increases with humidity. A critical missing piece of any such theory is a prediction of convective mass flux itself. Further complicating the matter, the GSRMs show that the population-mean buoyancy of convective updrafts declines with increasing humidity, contradicting the buoyancy calculations from idealized entraining plumes. We fill those gaps with an  entraining parcel model. This simple model shows how the convective mass flux response to humidity is a consequence of survival and dilution. In a humid environment, more parcels ascend rather than detrain, and these surviving parcels have become more diluted by entraining a greater mass of environmental air. Finally, we show that the transition to strong convection is accompanied by declining convective buoyancy in the parcel model. This is a consequence of convection becoming more diluted in humid environments.

Our paper is organized as follows. In Section \ref{sec:data} we describe the GSRM data as well as our calculation of convective mass flux. The GSRMs' precipitation response to humidity is documented in Section \ref{sec:prec}, and their mass flux response to humidity is documented in Section \ref{sec:massflux}. From there, our focus turns to understanding these results. In Section \ref{sec:buoyancy}, we show that the pickups in precipitation and mass flux are not associated with an increase in convective buoyancy. In Section \ref{sec:parcels}, we construct an entraining parcel model which helps us to explain the mass flux response to humidity. In Section \ref{sec:area}, we show how increases in updraft area fraction, rather than updraft intensity, are responsible for the mass flux response to humidity. In section \ref{sec:conclusion}, we discuss what this all means for our conceptual understanding of tropical convection, and in section \ref{sec:discussion} we discuss some of the limitations and opportunities presented by this study.

\section{Data}\label{sec:data}

This study relies upon simulations performed for phase two of the DYAMOND initiative, also known as DYAMOND Winter \cite{Stevens2019,Duras2021}. These global atmospheric simulations are run at an approximate horizontal resolution of 2 km to 5 km, sufficiently fine that deep convection may be permitted explicitly rather than parameterized. Each simulation is forced by historical sea surface temperatures and is run for approximately 40 days, from late January 2020 to the beginning of March 2020. Due to the computational expense of some of our calculations, we selected a limited set of five DYAMOND models for this study. These are (i) ARPEGE \cite{Roehrig2020}, (ii) NASA GEOS \cite{Putman2011}, (iii) SAM \cite{Khairoutdinov2022}, (iv) SHiELD \cite{Harris2020}, and (v) UM \cite{Walters2019} . For the sake of expediency, the specific selections were made on the basis of how readily we could apply a conservative regridding algorithm to the simulation outputs.

We have computed convective mass fluxes from the simulation outputs, regridded to a $1^\circ$ resolution to approximate the resolution of GCMs. To calculate mass fluxes, we employ instantaneous, three-dimensional model output at native GRSM resolution. Borrowing from past studies of convective mass flux \cite{Romps2010a,Williams2025}, we define convectively active air as any grid cell with a positive vertical velocity exceeding 1 m/s and cloud water mass fraction exceeding $1 \times 10^{-5}$ kg/kg. Convective mass flux is then calculated according to Eq. 1. Our specific procedure is to vertically interpolate the relevant fields to pressure levels, save the product  $w\rho$ in cells identified as convectively active and save zeros in all other cells, then apply a conservative (area-weighted) regridding algorithm. We explored using an alternative vertical pressure velocity threshold of -5 Pa/s and found it did not change our qualitative results. We do not consider convective downdrafts in this study. We similarly created mass-flux-weighted averages of the temperature and humidity of convectively active air and area-weighted averages of temperature and humidity in the non-convective environment, for which the calculations are more thoroughly described in Section \ref{sec:buoyancy}. All data reported in this study are conservatively regridded to a one-degree resolution. All data reported in this study are daily averages.

All of the analysis in this paper draws from data over the tropical oceans in a latitude band from $-10^\circ$ to $10^\circ$. Land is excluded so that differences between the terrestrial and oceanic environments will not confound our analysis. Column thermodynamic statistics, such as column humidity, are calculated on model output which has been regridded to a $1^\circ$ resolution in the horizontal and 25 hPa in the vertical. As an observational baseline, we also employ two reanalysis datasets, MERRA-2 and ERA-5 \cite{Gelaro2017,GlobalModelingandAssimilationOffice2015,Hersbach2020}. These reanalyses are matched to IMERG v7 Final Precipitation product, a multi-satellite-derived global precipitation data set, which we have regridded to a $1^\circ$ resolution \cite{Huffman2023}. We also use data from the Tropical Rainfall Measurement Mission (TRMM), taking advantage of its rain type classification algorithm, which separates precipitation into convective, stratiform, and shallow components \cite{Funk2013}.

\begin{figure}
    \centering
    \includegraphics[width=1.00\linewidth]{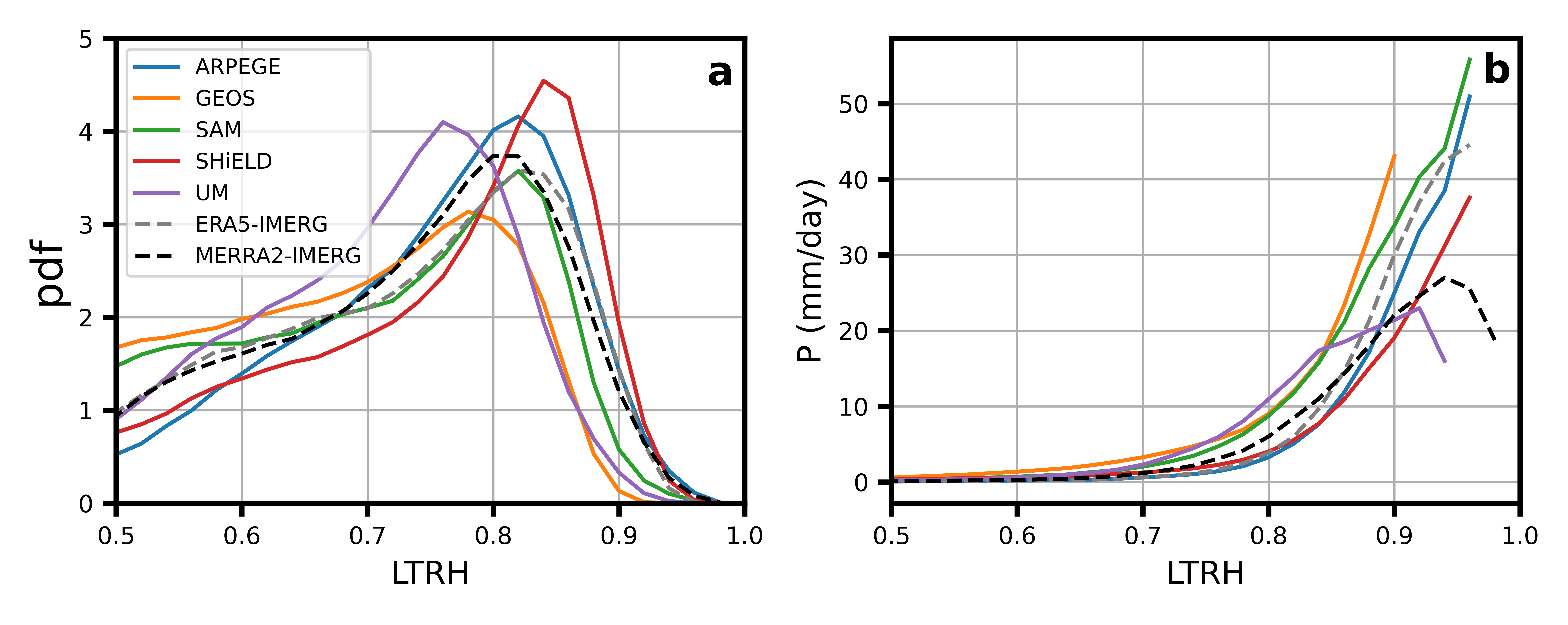}
    \caption{(a) Probability distribution function of $LTRH$ for the GSRMs and reanalysis. (b) Precipitation conditionally averaged over $LTRH$for the GSRMs and for IMERG precipitation (matched to reanalysis $LTRH$).}
    \label{fig:p_pickup}
\end{figure}

\section{Response of precipitation to environmental humidity}\label{sec:prec}

We are interested in how humidity influences convection, rather than the other way around. Therefore, we focus on humidity approximately below the freezing level, where convection initiates and most convective entrainment occurs. Our humidity measure of choice is Lower Tropospheric Relative Humidity (LTRH), the ratio of vertically integrated specific humidity ($q$) to vertically integrated saturation specific humidity ($q^*$) between 1000 hPa and 600 hPa:

\begin{linenomath*}
\begin{equation}
LTRH = \frac{ \int_{600\ hPa}^{1000\ hPa} q \,dp }{ \int_{600\ hPa}^{1000\ hPa} q^* \,dp }
\end{equation}
\end{linenomath*}

Convective mass flux is poorly observed at a global scale. Therefore, to provide a comparison between the GSRMs and observations, we shall first evaluate the relationship between mean precipitation and $LTRH$. To do so, we daily-average precipitation data into $LTRH$ bins, with a bin width of 0.02. Figure \ref{fig:p_pickup}a shows the probability distribution function of LTRH for each of the five models and the two reanalysis data sets. Each distribution has a wide left tail pronounced mode at at an $LTRH$ value between .75 and .85. The mode is conventionally thought to be near a ``critical point" in column humidity towards which the atmosphere relaxes by way of large-scale processes (if to the left of the critical point) or by convection (if to the right of the critical point) \cite{Peters2006,Neelin2008,Wolding2022}. We set the left bound of the figure at LTRH=0.5 in order to focus on the most humid environments. 

Figure \ref{fig:p_pickup}b shows the dependence of daily-average precipitation upon LTRH. The precipitation-humidity relationship is qualitatively similar between the five models and the observations. In each model, the precipitation increases nonlinearly with humidity. Models whose LTRH distributions have their mode at relatively low LTRH, such as UM and GEOS, tend to exhibit a pickup in precipitation at lower humidities. The SHiELD model, which has the greatest mode in its LTRH distribution, also exhibits the most ``delayed" pickup in precipitation. This corroborates the previously stated intuition that the pickup in precipitation is linked to the maximum in the humidity distribution.

\section{Response of mass flux to environmental humidity}\label{sec:massflux}

\subsection{Shift to deep inflow}

\begin{figure}
    \centering
    \includegraphics[width=1\linewidth]{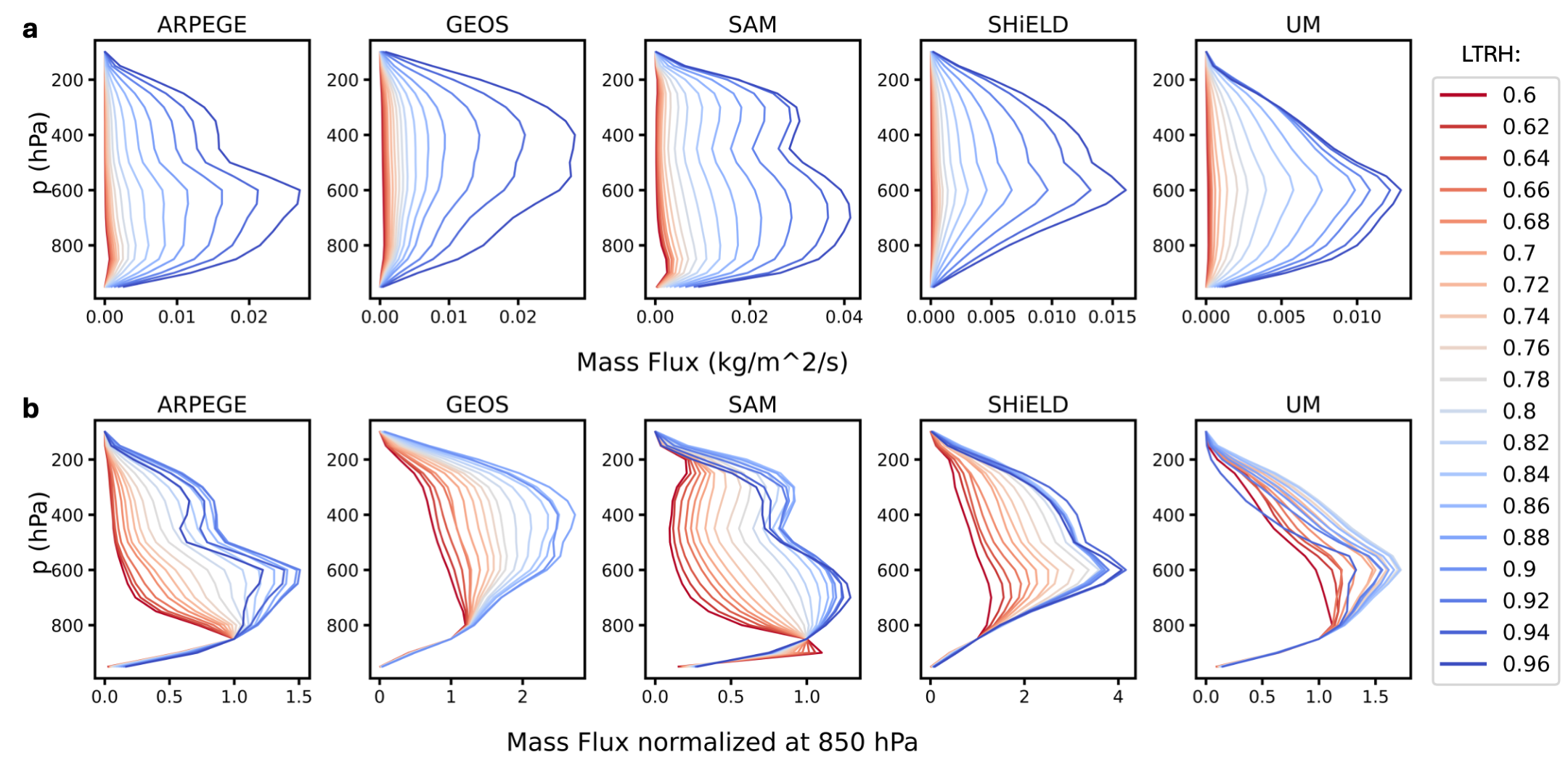}
    \caption{(a) GSRM convective mass flux conditionally averaged over $LTRH$. Each curve color denotes an $LTRH$ bin. (b) Normalized convective mass flux ($M/M_{850\ hPa}$) in the GSRMs, conditionally averaged over $LTRH$.}
    \label{fig:mprofs}
\end{figure}

Unlike surface precipitation, mass flux has a vertical dimension. As we shall see, not only does the magnitude of mass flux rise with increasing lower tropospheric humidity, but the shape of the mass flux profile shifts as well. The top row of panels in Fig. \ref{fig:mprofs} display the GSRMs' convective mass flux profiles, binned according to LTRH. These vertical profiles reveal a rich diversity of behaviors among the models. In all models, the convective mass flux increases dramatically with column humidity. However,  the horizontal axis scales are not the same for each model: in the most humid environments, there is an approximately three-fold difference in convective mass flux across the models. This spread is consistent with that for precipitation in Fig. \ref{fig:p_pickup}b. Furthermore, at elevated humidities, there are considerable differences in the vertical distribution of mass flux. Two models (SHiELD, UM) exhibit a peak in mass flux around 600 hPa, one model (GEOS) exhibits a peak in mass flux around 400 hPa, and two models (ARPEGE, SAM) exhibit a bimodal or nearly bimodal vertical structure in mass flux.

It is useful to know whether the mass flux increases or decreases with height in the lower free troposphere, and by how much, as these contain information about entrainment, detrainment, and the circulations that the clouds are embedded within. We have normalized each mass flux profile by dividing by its value at 850 hPa. These normalized mass fluxes are displayed in the lower panels of Fig. \ref{fig:mprofs}. Within the lower free troposphere, one pattern is consistent among the five models: At low humidity the normalized mass flux profile declines with height or is invariant ($M_{600} \leq M_{850}$). At elevated humidity, mass flux increases with height ($M_{600} > M_{850}$). The UM model is somewhat of an outlier, as it reaches maximum deep inflow at intermediate LTRH rather than the greatest LTRH. This may be related to UM's behavior in \ref{fig:p_pickup}b, where it exhibits the weakest pickup in precipitation among the five models. When the mass flux profile increases with height, the convection is sometimes said to be undergoing a ``deep inflow",. Deep inflow has been found in radar observations of tropical deep convection, and its presence indicates that more air is being entrained into convective clouds than detrained from them \cite{Schiro2018}. We adopt the same term to describe our conditionally-averaged mass flux profiles. The GSRMs exhibit a shift to deep inflow with increasing humidity.

This result from the GSRMs suggests that the existence or magnitude of deep inflow is not a universal feature of cumulus convection in the mid-troposphere. Rather, the shape of the mass flux profile depends upon the environment itself. The fractional rates of entrainment and/or detrainment systematically depend upon the environment. For the transition to deep inflow to occur, either entrainment must increase with LTRH, or detrainment must decrease, or both. This is consistent with \citeA{Becker2021}, who suggested that deep convection has a greater bulk entrainment rate in a humid environment. In a different vein, \citeA{Kuo2022} proposed that when convective buoyancy is organized into a region of large horizontal scale, the rising air may induce greater mass convergence below convective systems. If the horizontal scale of buoyancy is greater in humid environments, this would cause deep inflow in humid environments through a strictly dynamical mechanism. Another influence on the mass flux profile is geography, as different regions of the tropics have different mass flux profiles due to differences in surface temperature and large-scale dynamics \cite{Back2009}. However, the shift to deep inflow is robust to our choice of geography, appearing in both the Western and Eastern Pacific (See Fig. S1 in the Supporting Information). Furthermore, the shift to deep inflow is evident in the mass flux profiles previously reported in cases of simulated convective self-aggregation over a uniform surface temperature, suggesting that this is not a surface-temperature-driven phenomenon \cite{Igel2017}. 

\subsection{Pickup in mid-level convective mass flux}

\begin{figure}
    \centering
    \includegraphics[width=1.00\linewidth]{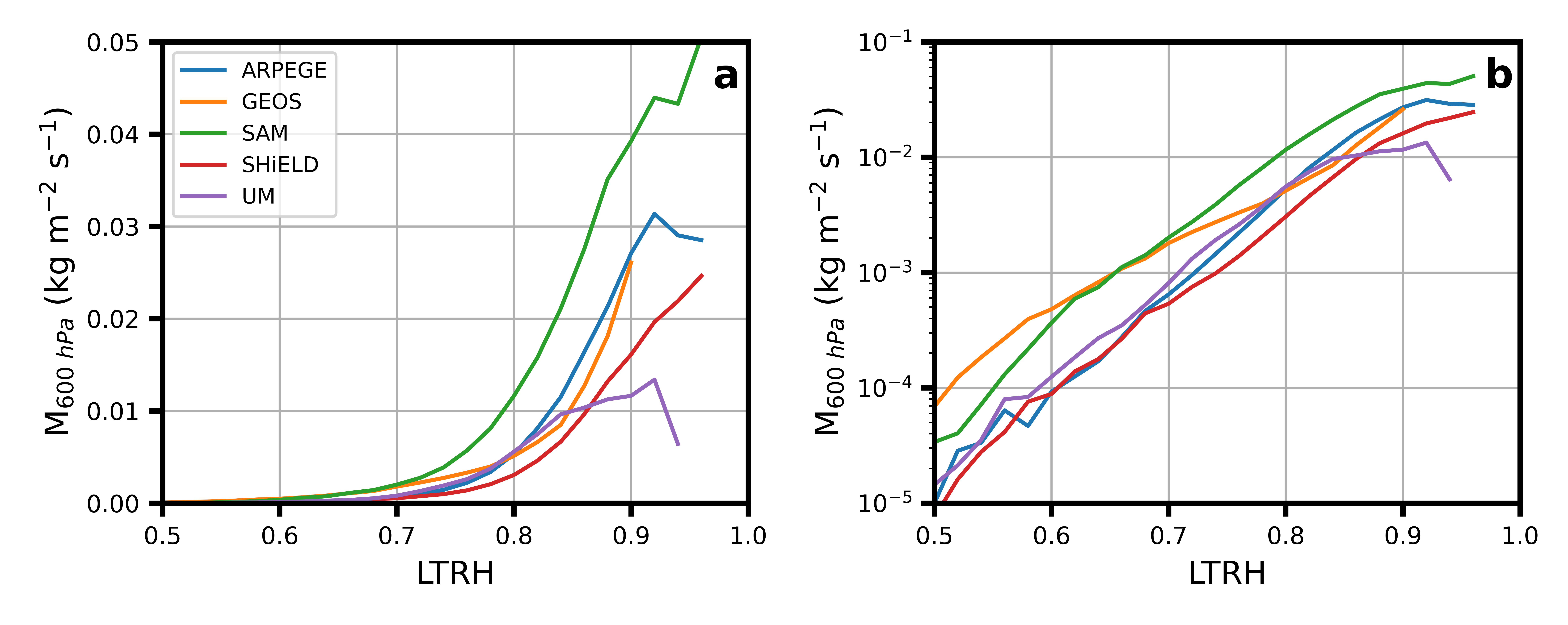}
    \caption{(a) GSRM convective mass flux at 600 hPa, conditionally averaged over $LTRH$. (b) As in (a), except with mass flux plotted on a logarithmic scale.}
    \label{fig:m_pickup}
\end{figure}

Now we will focus on the mass flux at 600 hPa ($M_{600}$), which we take to measure the magnitude of deep convection. We choose this vertical level because it is high enough that shallow convection will not be present, yet it is below the freezing level above which microphysical processes might introduce greater modeling uncertainty. Figure \ref{fig:m_pickup}a shows the relationship between $M_{600}$ and $LTRH$.  The five models exhibit a qualitatively consistent response in mid-tropospheric mass flux to LTRH. The pickup in $M_{600}$ broadly resembles the pickup in precipitation, albeit with a wider intermodel spread in relative magnitude.

We can clarify the nature of the mass flux pickup with LTRH by plotting mass flux on a logarithmic scale (Fig. \ref{fig:m_pickup}b). In this format, it becomes apparent that the pickup in mass flux does not occur at any particular value of humidity. Rather, $M_{600}$ has a quasi-exponential relationship to $LTRH$, and the pickup in $M_{600}$ is spread throughout much of the LTRH distribution. The $LTRH$-$M_{600}$ relationship spans several orders of magnitude in each model. For most models, the mass flux-humidity relationship becomes sub-exponential at the upper extreme in $LTRH$. This is consistent with past work on tropical precipitation which suggests that the relationship between mass flux and entraining buoyancy (which is a humidity-like metric) becomes linear above the critical point \cite{Ahmed2020}. However, \citeA{Wolding2024} reported an opposite result using a measure of vertically integrated lower-tropospheric buoyancy: precipitation increases \textit{more} rapidly on a log scale after the critical point.

\subsection{Two-layer analysis of mass flux}\label{sec:twolayer}

Why is deep convection more prevalent in humid environments? The interaction between humidity and convection runs in both directions, as convection may either moisten or dry the environment, and a humid environment usually supports stronger convection. In this paper, we are most interested in how humidity supports convection, for which there are two apparent causal pathways. First, a more humid boundary layer generally has greater moist static energy, so any undiluted parcels which ascend from there will be more buoyant. Second, when parcels do become diluted by entrainment, they have greater buoyancy if they entrain humid air rather than dry air. We shall demonstrate how LTRH contains information related to both of these processes. To do so, we construct two metrics. The first metric is related to the buoyancy of air undergoing undiluted ascent. It is calculated by taking the layer-average difference between the equivalent potential temperature $(\theta_{e}$) of air in the boundary layer ($900\ -\ 1000\ hPa$) and the saturation equivalent potential temperature ($\theta_{e}^*$)of air in the lower free troposphere ($600\ -\ 900\ hPa$):

\begin{linenomath*}
\begin{equation} \label{eq:undiluted}
\Delta\theta_{e,bl} = \frac{ \int_{900\ hPa}^{1000\ hPa} \theta_{e} \,dp }{100 hPa} - \frac{ \int_{600\ hPa}^{900\ hPa} \theta_{e}^* \,dp}{300 hPa}
\end{equation}
\end{linenomath*}

The second metric we construct is a measure of the subsaturation of the lower free troposphere. Entrainment weakens convection more in a dry environment than a humid environment.  This metric is calculated by taking the layer-average difference between equivalent potential temperature and saturation equivalent potential temperature in the lower free troposphere: 

\begin{linenomath*}
\begin{equation} \label{eq:subsat}
\Delta\theta_{e,ft} = \frac{ \int_{600\ hPa}^{900\ hPa} \theta_{e} \,dp }{300 hPa} - \frac{ \int_{600\ hPa}^{900\ hPa} \theta_{e}^* \,dp}{300 hPa}
\end{equation}
\end{linenomath*}

The integrals in eqs. \ref{eq:undiluted} and \ref{eq:subsat} are computed using the trapezoidal method on data which were vertically interpolated to a 25 hPa resolution.

\begin{figure}
    \centering
    \includegraphics[width=1\linewidth]{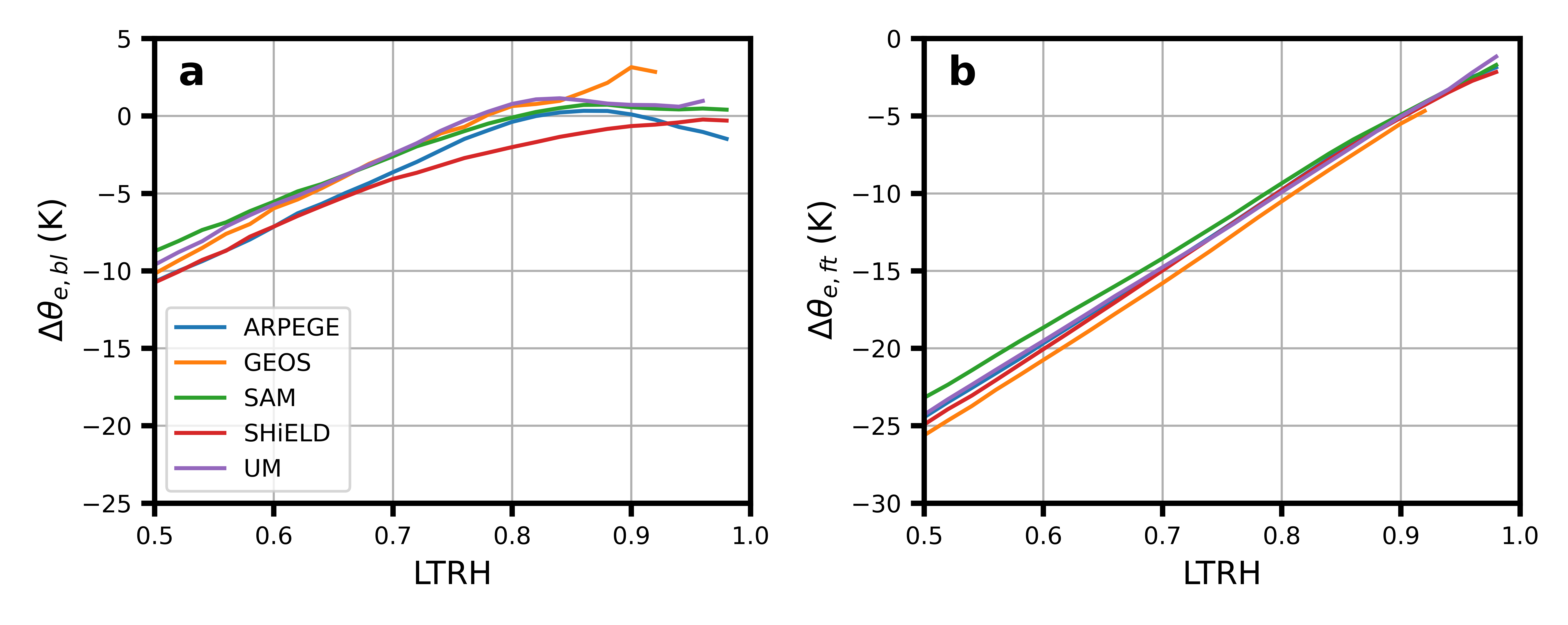}
    \caption{(a) $\Delta \theta_{e,bl}$: The difference between boundary-layer $\theta_{e}$ and free-troposphere $\theta_{e}^*$ in the GSRMs, conditionally averaged over $LTRH$. (b) $\Delta \theta_{e,ft}$: The difference between free-troposphere $\theta_{e}$ and free-troposphere $\theta_{e}^*$ in the GSRMs, conditionally averaged over $LTRH$. }
    \label{fig:dth_lth}
\end{figure}

Figure \ref{fig:dth_lth}a shows that $\Delta \theta_{e,bl}$ increases with LTRH, and Figure \ref{fig:dth_lth}b shows that $\Delta \theta_{e,ft}$ increases with LTRH. Therefore, $LTRH$ may increase $M_{600}$ by way of either undiluted instability or moist entrainment. However, if we narrow our attention to values of LTRH greater than 0.8, where the most dramatic rise in $M_{600}$ occurs (Fig. \ref{fig:m_pickup}), then the range of $\Delta \theta_{e,bl}$ is relatively small. In some models the trend in $\Delta \theta_{e,bl}$ is even neutral or declining $LTRH$. This suggests that the mass flux pickup at high humidities is primarily the result of free-troposphere entrainment. \citeA{Holloway2009} reached a similar conclusion by investigating the buoyancy of idealized entraining plumes. They found that variability in free-tropospheric humidity had a greater influence on plume buoyancy than did variability in boundary-layer humidity.

\begin{figure}
    \centering
    \includegraphics[width=1\linewidth]{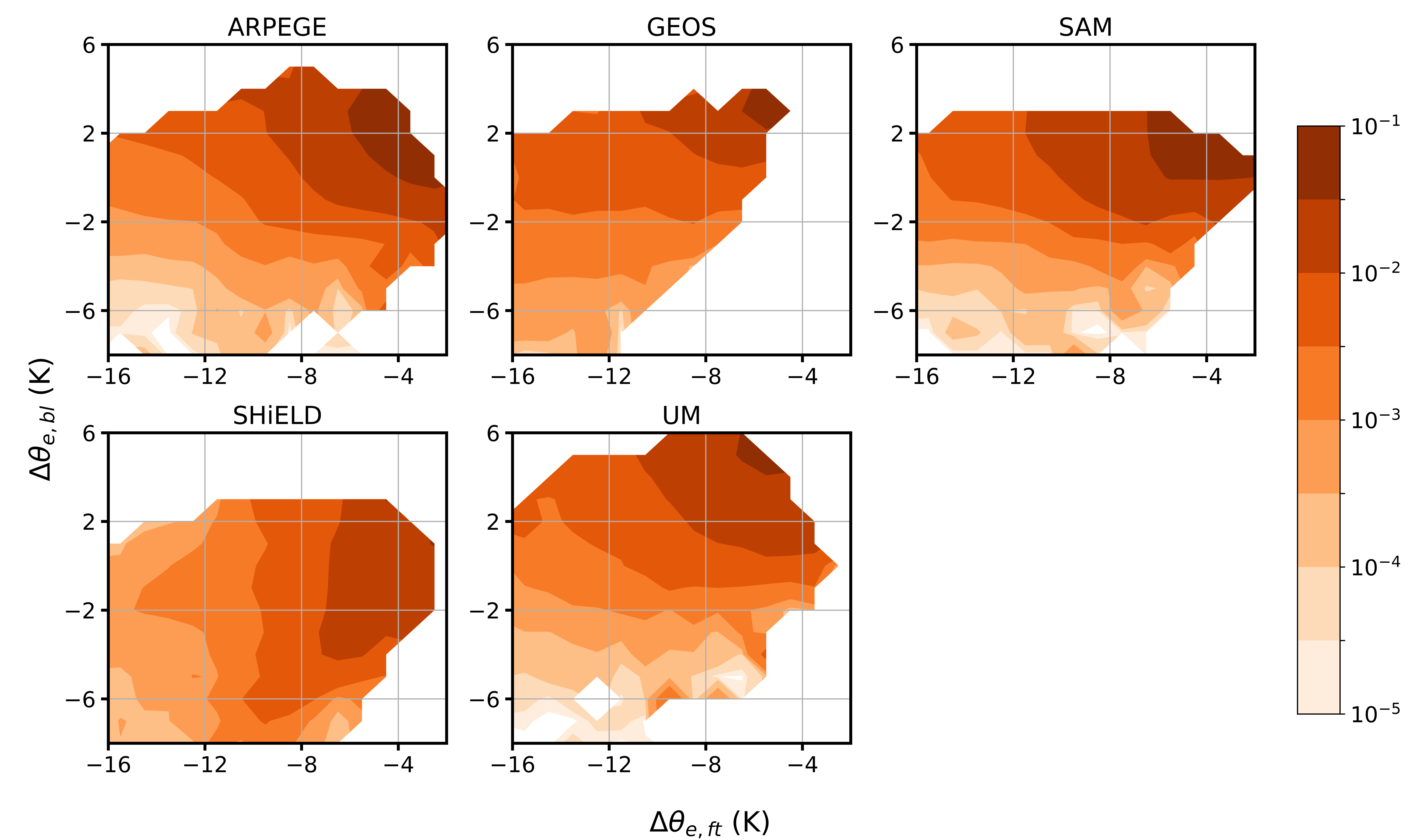}
    \caption{ Convective mass flux at 600 hPa, conditionally averaged over $\Delta \theta_{e,bl}$ and $\Delta \theta_{e,ft}$, for the five GSRMs in this study.}
    \label{fig:m_dth}
\end{figure}

We have concluded that the pickup in $M_{600}$ at the greatest values of LTRH is principally due to free-troposphere entrainment. However, $\Delta \theta_{e,bl}$ and $\Delta \theta_{e,ft}$ may nevertheless vary independently from one another. Therefore, it is useful to visualize the joint dependence of $M_{600}$ upon $\Delta \theta_{e,bl}$ and $\Delta \theta_{e,ft}$, as we have done in Fig. \ref{fig:m_dth}. With the exception of the SHiELD model, convective mass flux is sensitive to \textit{both} boundary layer energy and free-troposphere subsaturation. The tilt of the contours indicates the relative sensitivity of $M_{600}$ to a unit increase in $\Delta \theta_{e,bl}$ versus a unit increase in $\Delta \theta_{e,ft}$. $M_{600}$ is generally more sensitive to $\Delta \theta_{e,bl}$ than to $\Delta \theta_{e,ft}$. This is especially true at small $\Delta \theta_{e,bl}$. At large $\Delta \theta_{e,bl}$, the contours become more tilted, indicating that the influence of $\Delta \theta_{e,ft}$ is greater there. However, returning our attention to Fig. \ref{fig:dth_lth}, the range of $\Delta \theta_{e,ft}$ is considerably greater than the range of $\Delta \theta_{e,bl}$. There is a tradeoff between $M_{600}$'s greater unit sensitivity to $\Delta \theta_{e,bl}$ on the one hand and the greater variability in $\Delta \theta_{e,ft}$ on the other hand. This suggests that both quantities are important for governing the magnitude of mass flux.

The SHiELD model is a significant outlier in Fig. \ref{fig:m_dth}. Unlike the other four models, SHiELD's mass flux is much more sensitive to $\Delta \theta_{e,ft}$ than to $\Delta \theta_{e,bl}$. This is a surprising result, given that SHiELD's behavior is largely in line with that of the other models in Figs. \ref{fig:p_pickup}, \ref{fig:mprofs}, and \ref{fig:m_pickup}. This behavior is especially interesting when one considers that SHiELD and GEOS rely upon the same GFDL FV3 dynamical core. It may be rooted in differences in the sub-grid parameterizations. For example, \citeA{Abbot2024} reported that the intensity of resolved updrafts is remarkably sensitive to SHiELD's shallow convection parameterization.

\section{Decline in convective buoyancy with increasing humidity}\label{sec:buoyancy}

In the previous section, we reasoned that the response of convective mass flux to its environment is governed by the buoyancy of entraining plumes. This purported buoyancy-humidity relationship is a central theme in studies of deep convection. Traditional convection parameterizations often employ a ``quasi-equilibrium" closure in which cloud-base mass flux is assumed to be proportional to the potential energy available to convection, measured by the vertically integrated buoyancy of an entraining plume \cite{Arakawa1974,Betts1986,Zhang1995}. Separately, the pickup in tropical precipitation with humidity is often explained by reference to the fact that greater humidity causes greater buoyancy in a representative entraining plume, reasoning that an environment with greater plume buoyancy is more permissive of convection \cite{Holloway2009,Ahmed2018,Wolding2024}. In each of these cases, the buoyancy of a representative entraining plume is used to measure how hospitable the environment is to convection. This framework might lead us to na\"{\i}vely conclude that the buoyancy of \textit{actual} convective plumes increases with humidity. We believe that it will be instructive to investigate that na\"{\i}ve conclusion. As we shall see, the mean updraft buoyancy does not increase with humidity.

We wish to quantify how the buoyancy of convectively active air depends upon the humidity of the environment. To do so, we will measure buoyancy by the difference in virtual temperature between convection and its environment, $T_v'$. First, we calculate the instantaneous mass-flux-weighted virtual temperature $\overline{T_{v,c}}$ of convectively active air in the GSRMs:

\begin{linenomath*}
\begin{equation}
\overline{T_{v,c}} = \frac{\iint_{A} M\ T_{v,c}\ \,dA}{\iint_{A} M \,dA}
\end{equation}
\end{linenomath*}

where $T_{v,c}$ is the virtual temperature of convectively active air, as identified by the vertical velocity and cloud condensate thresholds previously described, and A is the total geographic area over which we are averaging. This computation is performed when coarsening the data from the native model resolution to the the one degree resolution used for analysis. We have chosen a mass-flux-weighted average over an area-weighted average so that we may faithfully compare our analysis in this section to the results of the parcel model in the next section, which does not predict convective area. Our qualitative results are nevertheless robust to the choice of either area or mass flux weighting for $T_{v,c}$ (Fig. S2 in the Supporting Information). When handling the data at native resolution, we neglect horizontal variations in density, so mass flux scales with the multiple of vertical velocity and grid box area.

Separately, we calculate an instantaneous area-weighted average of virtual temperature $<T_{v,e}>$ of convectively inactive air:

\begin{linenomath*}
\begin{equation}
<T_{v,e}> = \frac{\iint_{A} \sigma_{e}\ T_{v,e}\ \,dA}{\iint_{A} \sigma_{e} \,dA}
\end{equation}
\end{linenomath*}

where $T_{v,e}$ is the virtual temperature of convectively inactive air and $\sigma_{e}$ is the area fraction of non-convective air.

To characterize the mean buoyancy of the population of convective updrafts, we shall use the average virtual temperature perturbation of convectively active air with respect to its environment:

\begin{linenomath*}
\begin{equation}
T_v' = \overline{T_{v,c}} - <T_{v,e}>
\end{equation}
\end{linenomath*}

\begin{figure}
    \centering
    \includegraphics[width=0.5\linewidth]{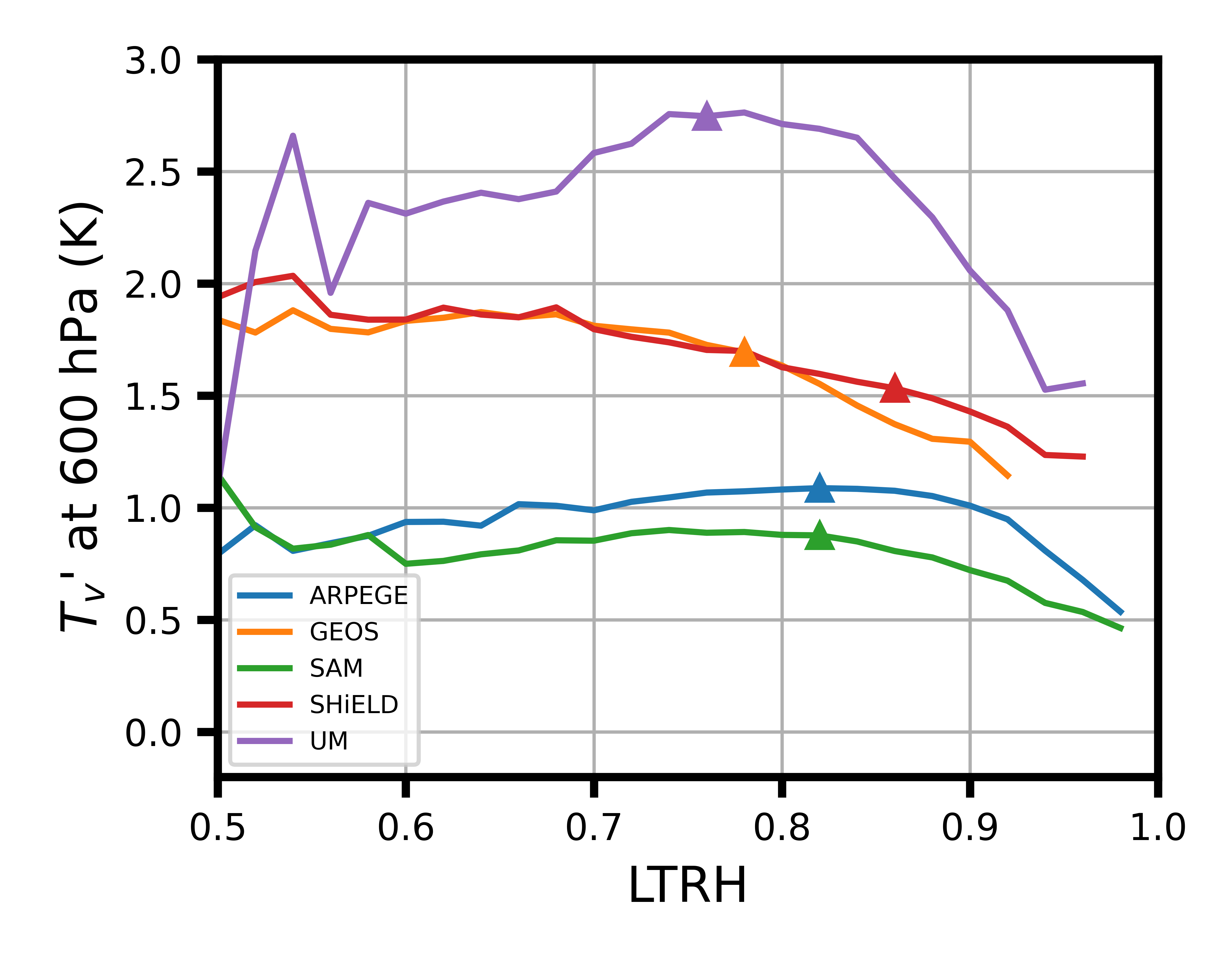}
    \caption{Mean updraft buoyancy, $T_v'$, as measured by the difference between the virtual temperature of convectively active air and the virtual temperature of the environment at 600 hPa in the GSRMs. The triangles indicate the modal LTRH.}
    \label{fig:tvprime_gsrms}
\end{figure}

If it is true that the pickups in precipitation and mass flux with humidity are due to greater convective buoyancy, then $T_v'$ should increase with $LTRH$. We then take a daily average of $T_v'$ and bin it according to $LTRH$, weighting by mass flux as we did above for $T_{v,c}$ in the one-degree grid boxes. Figure \ref{fig:tvprime_gsrms} shows the conditional average of $600\ hPa$ $T_v'$ on $LTRH$ for the five GSRMs in this study. To a first approximation, $T_v'$ changes fairly little with increasing $LTRH$. Furthermore, all five models show that $T_v'$ $declines$ at values of $LTRH$ above the mode of the $LTRH$ distribution, denoted by the triangle marks in Fig. \ref{fig:tvprime_gsrms}. In deep convective regimes, greater humidity is associated with \textit{reduced} convective buoyancy.  This is a surprising result. In most of the GSRMs, this decline in buoyancy is nevertheless small. To first order, the population-mean buoyancy of updrafts is invariant with respect to $LTRH$. We shall try to better understand this behavior in the next section.

\section{A parcel model for mass flux}\label{sec:parcels}

In order to better understand the physical basis of the GSRMs' mass flux response to humidity, we implement a bulk mass flux model. Our goal  is to articulate a minimum recipe for the following three responses to increasing humidity: (1) the transition to convective deep inflow, (2) the quasi-exponential pickup in mid-level mass flux, and (3) the decline in mean convective buoyancy. The equations for our model are identical to those for a bulk plume model \cite{Singh2013,Romps2014}. However, we refer to it as a \textit{parcel model}, imagining it to describe a population of rising parcels rather than a singular, homogeneous plume. The following equations govern mass, moist static energy, and specific humidity in the parcel model, respectively:

\begin{linenomath*}
\begin{equation} \label{eq:parcel_m}
\frac{d}{dz}[M] = \epsilon M - \delta M
\end{equation}
\end{linenomath*}

\begin{linenomath*}
\begin{equation} \label{eq:parcel_mse}
\frac{d}{dz}[Mh_p] = \epsilon M h_e - \delta M h_p
\end{equation}
\end{linenomath*}

\begin{linenomath*}
\begin{equation} \label{eq:parcel_q}
\frac{d}{dz}[Mq_p] = \epsilon M q_e - \delta M q_p - M c
\end{equation}
\end{linenomath*}

where \textit{M} is the mass flux of a stream of rising parcels in units of $kg\ m^{-2}\ s^{-1}$; $\epsilon$ and $\delta$ represent the fractional entrainment and detrainment rates, respectively, in $m^{-1}$; $h_p$ and $h_e$ represent the moist static energies of the parcels and environment, respectively, in $J/kg$; $q_p$ and $q_e$ represent the specific humidities of the parcels and environment, respectively, in $kg/kg$; and $c$ is a net sink in specific humidity due to condensation, in $1/m$. The moist static energy $h$ is defined as:

\begin{linenomath*}
\begin{equation}
h = c_pT + Lq + gz
\end{equation}
\end{linenomath*}

where $c_p$ is the specific heat capacity of air at constant pressure, T is the temperature, $L$ is the latent heat of vaporization, and $g$ is the acceleration due to gravity. 

Parcels rise in a dry-adiabatic fashion if they are subsaturated, and they rise in a moist-adiabatic fashion if saturated. During moist-adiabatic ascent, condensed water is immediately evacuated from the cloud, and we neglect the effect of this mass sink on mass flux. In the case of saturation, the equation for specific humidity is redundant with the equation for moist static energy. We also neglect the latent heat of fusion, as we are primarily interested in explaining phenomena at or below the freezing level.

The model employs buoyancy sorting to determine the rate of detrainment \cite{Raymond1986,Bretherton2004a}. Parcels cease their ascent and fully detrain if their virtual temperature is less than that of the environment. This instantaneous detrainment is accomplished by setting $\delta = 1/\Delta z$ and $\epsilon = 0$, where $\Delta z$ is the vertical resolution of the model. The entrainment rate is prescribed, as one of the purposes of this model is to explore how different treatments of entrainment affect convective mass flux.

The parcel model employs daily-average temperature and humidity data from MERRA-2 for the full year 2020. As with the GSRMs, we limit the domain to tropical oceans within $10^\circ$ of the equator. These data are horizontally interpolated to a one-degree resolution, and the vertical grid is interpolated to the 100m resolution of the parcel model. All parcels are launched from an altitude of 300 m, which is near the lowest level in the MERRA-2 data (at 1000 hPa). All parcels are initialized with a unit mass flux. To ensure that parcels have sufficient buoyancy and energy to ascend through the boundary layer, the parcel's initial temperature is set to $0.5 K$ warmer than its environment, and the relative humidity is fixed at the value in MERRA-2. We shall see that this rudimentary closure is sufficient to replicate the GSRMs' qualitative behavior. 

\subsection{Transition to deep inflow in the parcel model}

Our first goal is to identify a minimum recipe for the transition to convective deep inflow with increasing humidity. In the parcel model, our minimum recipe has two key ingredients: (a) entrainment and (b) detrainment by buoyancy sorting. 

\begin{figure}
    \centering
    \includegraphics[width=0.5\linewidth]{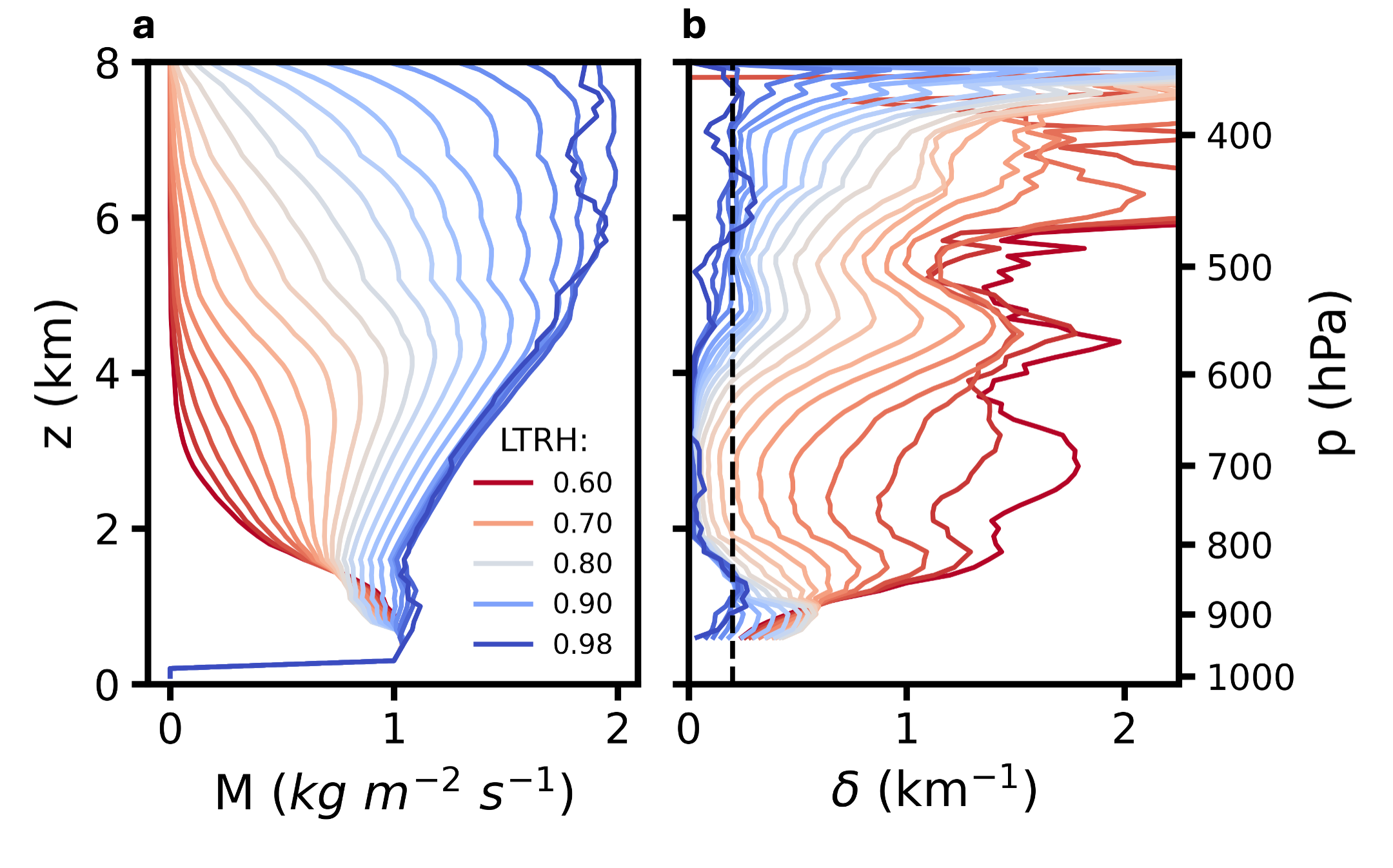}
    \caption{(a) Convective mass flux in the parcel model with a fixed entrainment rate of $\epsilon = 0.2 km^{-1}$ , binned by MERRA-2 $LTRH$. Each color represents a value of $LTRH$. The dashed black line indicates the prescribed fixed entrainment rate of $\epsilon=0.2\ km^{-1}$. (b) Fractional detrainment in the parcel model with a fixed entrainment rate of $\epsilon = 0.2 km^{-1}$, binned by $LTRH$. The dashed line marks the value of the entrainment rate. A 400 m moving average is applied to each detrainment profile in order to render it more interpretable.}
    \label{fig:profs_eps}
\end{figure}

For our initial set of parcel simulations, we prescribe a fixed fractional entrainment rate $\epsilon$ in the model.\footnote{One may worry that a fixed fractional entrainment rate is inappropriate if the wind shear or lifting condensation level (LCL) is strongly correlated with humidity \cite{Peters2022,Mulholland2021}. Figure S3 in the Supporting Information shows that vertical wind shear and LCL tend to be small and have relatively weak dependence on $LTRH$, so we choose to neglect these effects. The LCL was calculated using the \citeA{Romps2017} formula.} The model is run on daily-average MERRA-2 profiles, which are then binned according to $LTRH$. Each bin comprises many different parcel calculations due to within-bin variability in temperature and water vapor. Figure \ref{fig:profs_eps}a shows the bin-average mass flux profiles for a fractional entrainment rate of $\epsilon = 2 \times 10^{-4}\ m^{-1}$. Between 1 km and 6 km there is a shift from net detrainment ($\frac{dM}{dz} < 0$) at low humidity to net entrainment ($\frac{dM}{dz} > 0$) at high humidity. This is qualitatively similar to the mass flux profiles simulated by the five GSRMs we examined (Fig. \ref{fig:mprofs}). Therefore, the parcel model may explain the GSRMs' shift to deep inflow.

How is the shift to deep inflow accomplished in the parcel model? All parcels in the $\epsilon = 2 \times 10^{-4}\ m^{-1}$ simulation have identical rates of fractional entrainment. Therefore, variability in $\frac{dM}{dz}$ must be explained by the fractional detrainment rate $\delta$. Figure \ref{fig:profs_eps}b shows the $LTRH$-binned profiles of  $\delta$ for the parcel model when executed with $\epsilon = 2 \times 10^{-4}\ m^{-1}$. As $LTRH$ increases, fractional detrainment declines. Recall that the parcels fully detrain if and only if they are negatively buoyant. In a humid environment, entrainment is less likely to reduce the buoyancy of a parcel below zero, so an entraining parcel is more likely to survive its ascent. The average detrainment rate is consequently smaller in a humid environment than in a dry envronment. At a sufficiently large value of $LTRH$, $\delta < \epsilon$ in the lower free troposphere so that, per Eq. \ref{eq:parcel_m}, $\frac{dM}{dz} > 0$. This is the origin of the ``deep inflow" mass flux profile in the parcel model.

\subsection{Heterogeneous entrainment and the pickup in midlevel convective mass flux}

\begin{figure}
    \centering
    \includegraphics[width=1\linewidth]{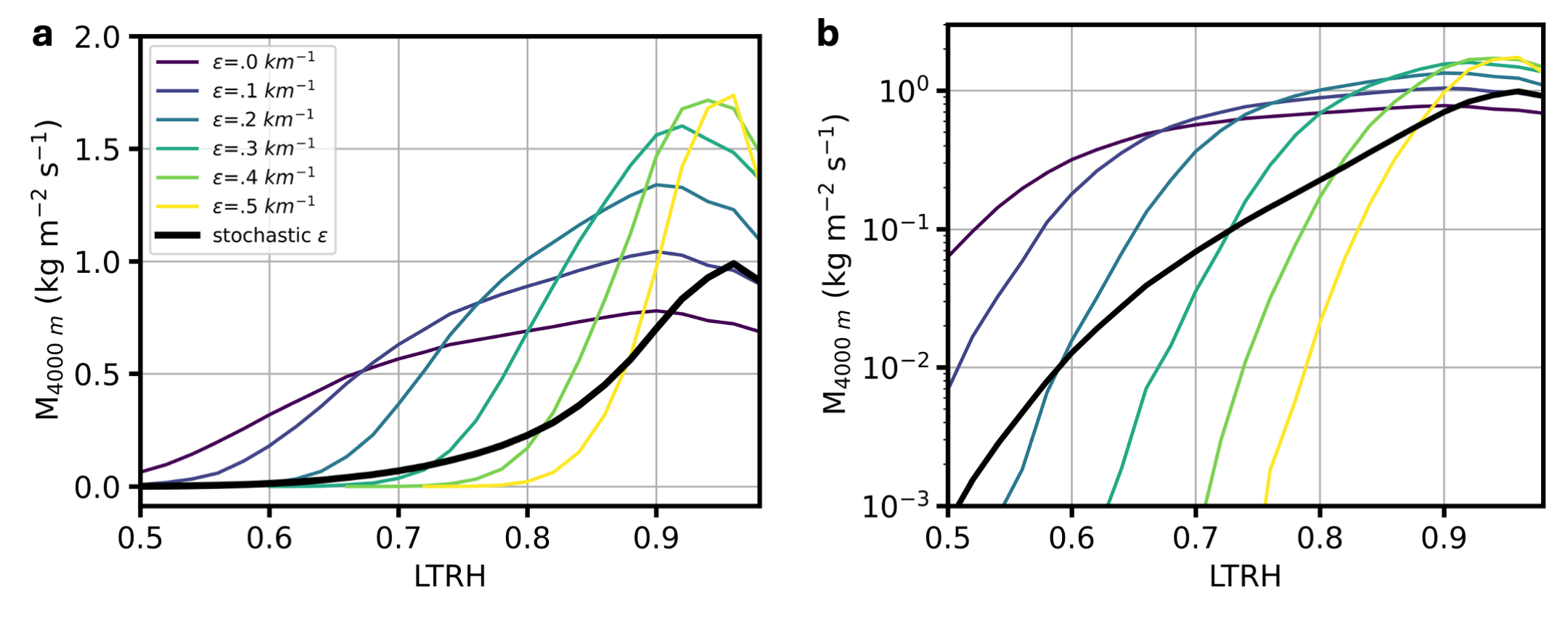}
    \caption{Parcel-model mass flux at $4000\ m$, conditionally averaged on MERRA-2 $LTRH$. (a) Linear scale. (b) Log scale.}
    \label{fig:m_pickup_spm}
\end{figure}

We shall now examine whether the parcel model can replicate the pickup in mid-level mass flux found in the GSRMs (Fig. \ref{fig:m_pickup}). The colored curves in Fig. \ref{fig:m_pickup_spm}a show how the parcel mass flux at 4000 m depends on $LTRH$ for several fixed rates of fractional entrainment. To avoid using interpolations, we have use the 4000 m model level of the parcel model to represent the mid levels, in place of the 600 hPa level we used in the GSRMs. At each rate of fractional entrainment in Fig. \ref{fig:m_pickup_spm}a, the parcel model fails to replicate the GSRMs' qualitative behavior. Below a minimum value of $LTRH$, there is no mass flux at all. Above it, mass flux increases in a nearly linear manner as a greater share of parcels survive its ascent (rather than detrain). This is qualitatively dissimilar to the GSRMs, where each humidity bin has some mass flux present in the mid-troposphere and the relationship between humidity and mass flux is quasi-exponential (Fig. \ref{fig:m_pickup}). In order to replicate the GSRMs' qualitative behavior, we need to add a little more complexity to our minimum recipe for mass flux. To do so, we shall discard the assumption of a fixed entrainment rate in favor of heterogeneous entrainment.

Aircraft-based observations have shown that energy and water are not distributed homogeneously within cumulus clouds \cite{Warner1955,Paluch1979,Blyth1988}. Convective clouds are internally heterogeneous. One possible explanation for this is that the parcels within a cloud have different entrainment rates from one another \cite{Romps2010,Suselj2014}. Similarly, \citeA{Morrison2020} suggested that rising thermals of air cause environmental air to penetrate the cloud below them, thus increasing the effective entrainment rate for successive thermals. Furthermore, the population of convective clouds is heterogeneous, spanning a range of sizes. Traditionally, narrower convective elements are thought to have greater rates of fractional entrainment \cite{Morton1956,Lecoanet2019}. More recently, \citeA{Kuo2022} showed that the large-scale rate of entrainment may also depend on both the vertical structure and horizontal scale of buoyancy perturbations.  However, the parcel model is too simple to directly simulate the complex cloud-scale dynamics which might give rise to heterogeneous entrainment. Instead, we will emulate those dynamics by implementing stochastic entrainment in the parcel model.

Our stochastic treatment of entrainment closely follows \citeA{Romps2010} and \citeA{Romps2016}. We set entrainment equal to zero except during discrete entrainment events which occur as a Poisson process with average return distance $\alpha = 500\ m$. This distance was chosen based on discussion in \citeA{Romps2016}. That study estimated a parameter value of 500 m for deep convection based on exponential decay of the undiluted mass flux with height in large-eddy simulations of deep convective clouds \cite{Romps2010a}. Since our treatment of stochastic entrainment is based in part on a large-eddy simulation result, it is informed by the complex cloud dynamics in those simulations. The fractional mass of air entrained, relative to the mass of a parcel, is drawn from an exponential distribution with mean $\mu = 0.25$. This is the value used in the \citeA{Romps2016} convection scheme. The expected fractional entrainment rate for any parcel is $\epsilon = \mu / \alpha = 0.5\ km^{-1}$. Individual parcel paths may have a greater or lesser value. This value of $\epsilon$ is consistent with that found in high-resolution simulations of deep convection \cite{Romps2010,Hannah2017}. Each entrainment event occurs over a distance of $\Delta z_\epsilon = 200 m$. 

\begin{figure}
    \centering
    \includegraphics[width=0.75\linewidth]{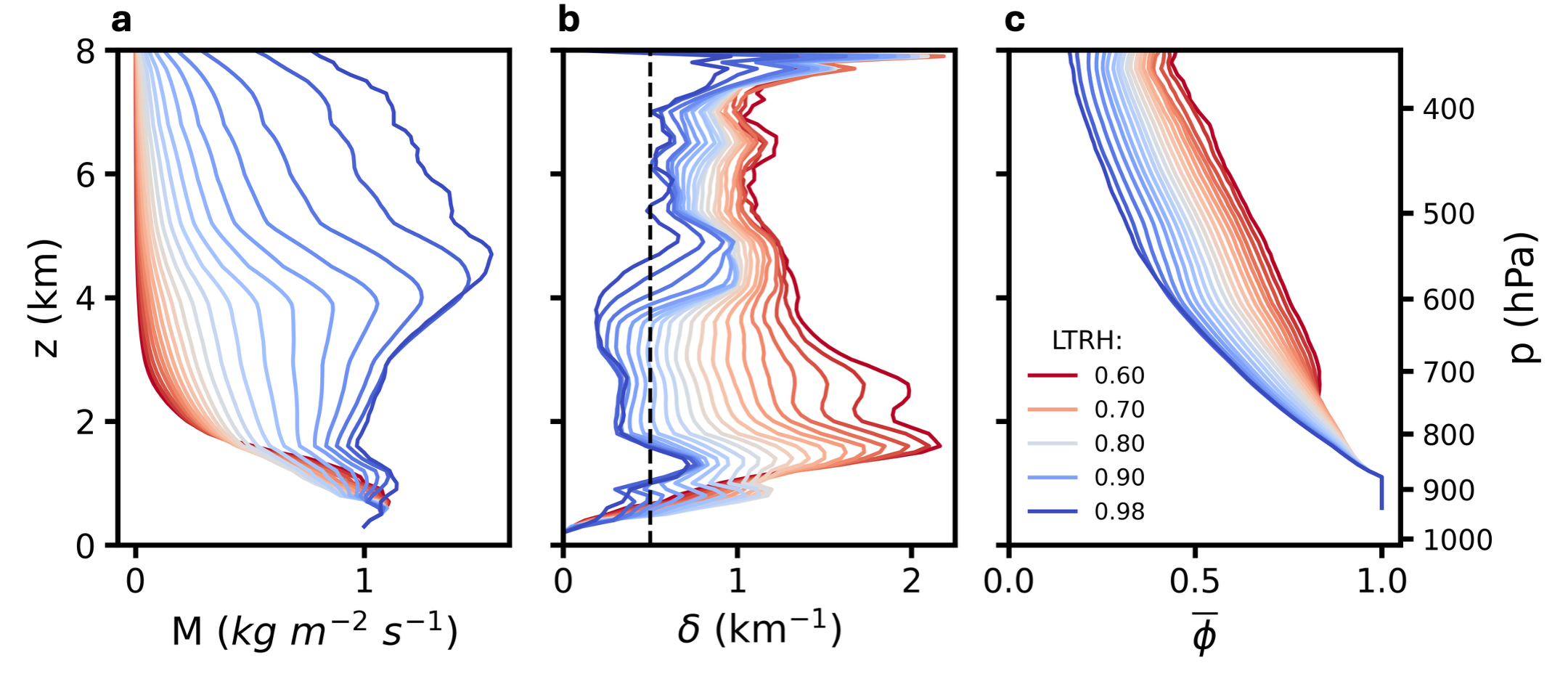}
    \caption{(a) Convective mass flux in the parcel model with a stochastic entrainment rate, binned by MERRA-2 $LTRH$. Each color represents a value of $LTRH$. (b) Fractional detrainment in the parcel model with a stochastic entrainment rate, binned by $LTRH$. A 400 m moving average is applied to each detrainment profile in order to render it more interpretable. The dashed black line indicates the average entrainment rate of $\epsilon=0.5\ km^{-1}$. (c) Average purity of convective mass flux in the parcel model.}
    \label{fig:profs_spm}
\end{figure}

We imagine that each one-degree MERRA-2 grid box contains parcels with all possible entrainment histories. To calculate the bulk properties of these parcels, we rely on the Monte Carlo method. For each daily-average column over the tropical ocean ($\pm 10^\circ$), we simulate 50 parcel trajectories. We then bin all of the model results according to humidity. We exclude any $LTRH$ bins with less than 100 occurrences in our data set. Thus the minimum number of simulations in a $LTRH$ bin is 5 000. We present the stochastic parcel mass flux profiles in Fig. \ref{fig:profs_spm}a and detrainment profiles in Fig. \ref{fig:profs_spm}b. The stochastic parcel mass flux profiles appear similar to those of the GSRMs in the free troposphere. As with fixed entrainment, there is a shift from to deep inflow below 4 km, which can be explained by a varying detrainment rate. In humid environments, mass flux declines with altitude above 4000 m, which is similar to several of the GSRMs but may be a consequence of our choice to neglect freezing and ice processes in the parcel model. 

How well do the stochastic parcels emulate the pickup in midlevel convective mass flux? The black curve in Fig. \ref{fig:m_pickup_spm}a shows that the stochastic parcel model exhibits a similar nonlinear pickup to what we found for the GSRMs. The similarity between the stochastic parcels and the GSRMs is even more striking when mass flux is plotted on a log scale (Fig. \ref{fig:m_pickup_spm}b). The stochastic parcels's mass flux pickup is quasi-exponential, much like that of the GSRMs (Fig. \ref{fig:m_pickup}b). In contrast, the fixed-entrainment mass fluxes in Fig. \ref{fig:m_pickup_spm}b pick up too quickly at low humidities and too slowly at high humidities. 

The stochastic parcels are also successful in replicating the \textit{magnitude} of the mass flux pickup. In both the GSRMs and in the stochastic parcel model, mass flux increases by approximately two orders of magnitude between $LTRH = 0.60$ and $LTRH = 0.90$. However, this is subject to tuning: The magnitude of the pickup is sensitive to the initial temperature of the parcels, the precise altitude the parcels are launched from, and the values of the entrainment parameters each influence the magnitude of the pickup. Without any information to better inform our choices in the parcel model, we simply used the entrainment parameters discussed for deep convection in \citeA{Romps2016}, a launch altitude at the near-surface level in MERRA-2, and a minimal buoyancy perturbation to allow parcels to rise from the boundary layer.

Why does the stochastic parcel model replicate the quasi-exponential relationship between mass flux and humidity, as found in the GSRMs? To answer this question, we can decompose the mass flux at any given level into two components: the survival fraction $f_{s}$, which denotes the fraction of parcels launched at the 300 m initialization level and ascended to that level without detraining, and the mean mass flux of surviving parcel trajectories $\overline{M_{s}}$:

\begin{linenomath*}
\begin{equation}
M = f_{s} \overline{M_{s}} 
\end{equation}
\end{linenomath*}

\begin{figure}
    \centering
    \includegraphics[width=0.75\linewidth]{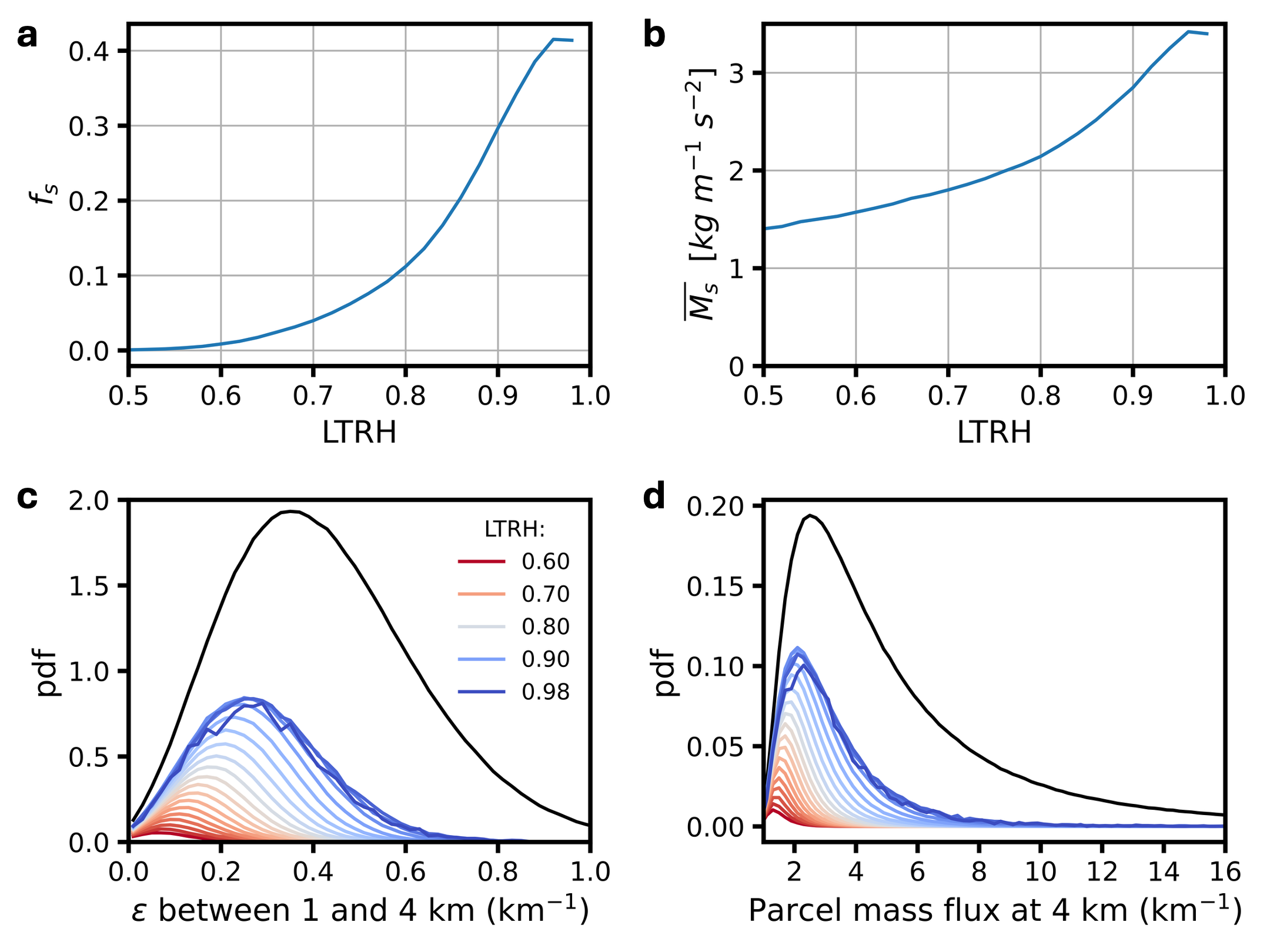}
    \caption{(a) Survival fraction in the stochastic parcel model, which denotes the fraction of stochastic parcel trajectories which are non-detraining up to 4000 m. (b) The mean mass flux of surviving parcel trajectories at 4000 m. (c) Distribution of stochastic parcel model entrainment rates between 1000 and 4000 m for varying $LTRH$ (colored lines) and for hypothetical non-detraining parcels (black line). (d) Distribution of stochastic parcel model mass fluxes at 4000 m for varying $LTRH$ (colored lines) and for hypothetical non-detraining parcels (black line).}
    \label{fig:distributions}
\end{figure}

Figures \ref{fig:distributions}a and Fig. \ref{fig:distributions}b show $f_{s}$ and $\overline{M_{s}}$ at 4000 m, plotted against $LTRH$. Both $f_s$ and $\overline{M_{s}}$ increase nonlinearly with $LTRH$. However, the mean mass flux of surviving trajectories $\overline{M_{s}}$ increases by little more than two-fold over the wide range of $LTRH$ plotted, while the pickup in $f_s$ appears to span at least two orders of magnitude. The pickup in mid-level mass flux is thus dominated by a pickup in rate at which parcels survive ascent rather than detraining.

What explains the nonlinear relationship between $LTRH$ and the survival fraction? The parcel model is designed such that the modal parcel has an entrainment rate which is too large for the parcel to remain buoyant in most environments. This can be seen in Fig. \ref{fig:distributions}c, which shows the distribution of entrainment rates over a 3km distance if no parcel detrains (black curve) alongside the realized distribution of entrainment rates among surviving parcels (colored curves). The total integral of any colored curve in Fig. \ref{fig:distributions}c is equal to the survival fraction $f_s$. The vertical difference between the black curve and any colored curve represents all the parcels that have detrained at that value of $\epsilon$. Except at the greatest humidities, an overwhelming share of parcels do not survive at the modal entrainment rate (about $0.4\ km^{-1}$). Thus, there is a greater number of parcels with slightly too much entrainment to survive than there is of parcels with just the right amount of entrainment. A small increase in humidity causes a disproportionate increase in the number of parcels which survive.

The average mass flux of surviving parcels is greater in humid environments (Fig. \ref{fig:distributions}b). This is because in humidity allows parcels to accommodate greater entrainment without becoming negatively buoyant and detraining. This causes the \textit{average} surviving parcel to have a greater entrainment rate. We refer to this effect as \textit{dilution}, as the greater mass flux reflects the fact that surviving parcels are more diluted by having entrained environmental air. Figure \ref{fig:distributions}c similarly shows that the surviving parcels become more diluted with increasing humidity, as the distribution of entrainment rates among survivors shifts to the right with increasing $LTRH$. Figure \ref{fig:distributions}d shows the distributions of mass fluxes for surviving parcels. At greater $LTRH$, more parcels with relatively large mass fluxes survive. Thus, the average surviving parcel's mass flux increases. Survival and dilution are not separate mechanisms but rather two consequences of the same process. The survival fraction is generally set by how diluted a parcel may become without detraining.

\subsection{The dilution rate}\label{sec:dilution}

We find it helpful to draw a distinction between entrainment and dilution \cite{Hannah2017}. Entrainment is a parcel-scale process in which convectively inactive air is incorporated into convectively active air. Dilution is the extent to which convectively active air at any given level was entrained from the environment during ascent, as opposed to having ascended from the cloud base. Dilution is a backwards-looking description of only the surviving parcels. Here we wish to quantify the dilution rate, or the average rate at which environmental air has been incorporated those parcels which has survived. In practice, this is nearly equivalent to the ``entrainment" rate used in a bulk plume model, as it quantifies how the thermodynamic properties of the surviving convection are influenced by entrainment \cite{Arakawa1974,Singh2013,Romps2014}. 

To quantify the dilution rate, we first define the purity $\phi_p$ of an individual parcel as:

\begin{linenomath*}
\begin{equation} \label{eq:purity_eq}
\phi_p =  \frac{M_b}{M}
\end{equation}
\end{linenomath*}

where $M_b$ is the mass flux at the level of the cloud base which we take as $1000\ m$ for simplicity. In numerical simulations, purity is ordinarily quantified using a tracer \cite{Romps2010,Romps2010a}. We take advantage of the parcel model's strict buoyancy sorting, which allows us to interpret each parcel's mass flux as a measure of the amount of air entrained. According to Eq. \ref{eq:purity_eq}, a parcel's purity declines with each entrainment event once it has ascended past the 1 km level. 

Figure \ref{fig:profs_spm}c shows the mass-flux-weighted average purity profiles in the stochastic parcel model. Purity generally declines with height, and it declines with increasing LTRH. To understand the latter behavior, we must consider how different parcels react to different environments. Due to heterogeneous entrainment, ascending parcels have different purities from one another. Detrainment is also heterogeneous, as each parcel ends its ascent according to its own moist static energy. The most impure parcels usually have the least moist static energy and thus the least buoyancy. Impure parcels are thus more likely to detrain. Detrainment consequently makes the surviving convection more concentrated (i.e., less diluted) on average. Entrainment makes convectively active air more diluted on average. To quantify the net effect of detrainment and entrainment on purity, we calculate an average dilution rate $\lambda$:

\begin{figure}
    \centering
    \includegraphics[width=.5\linewidth]{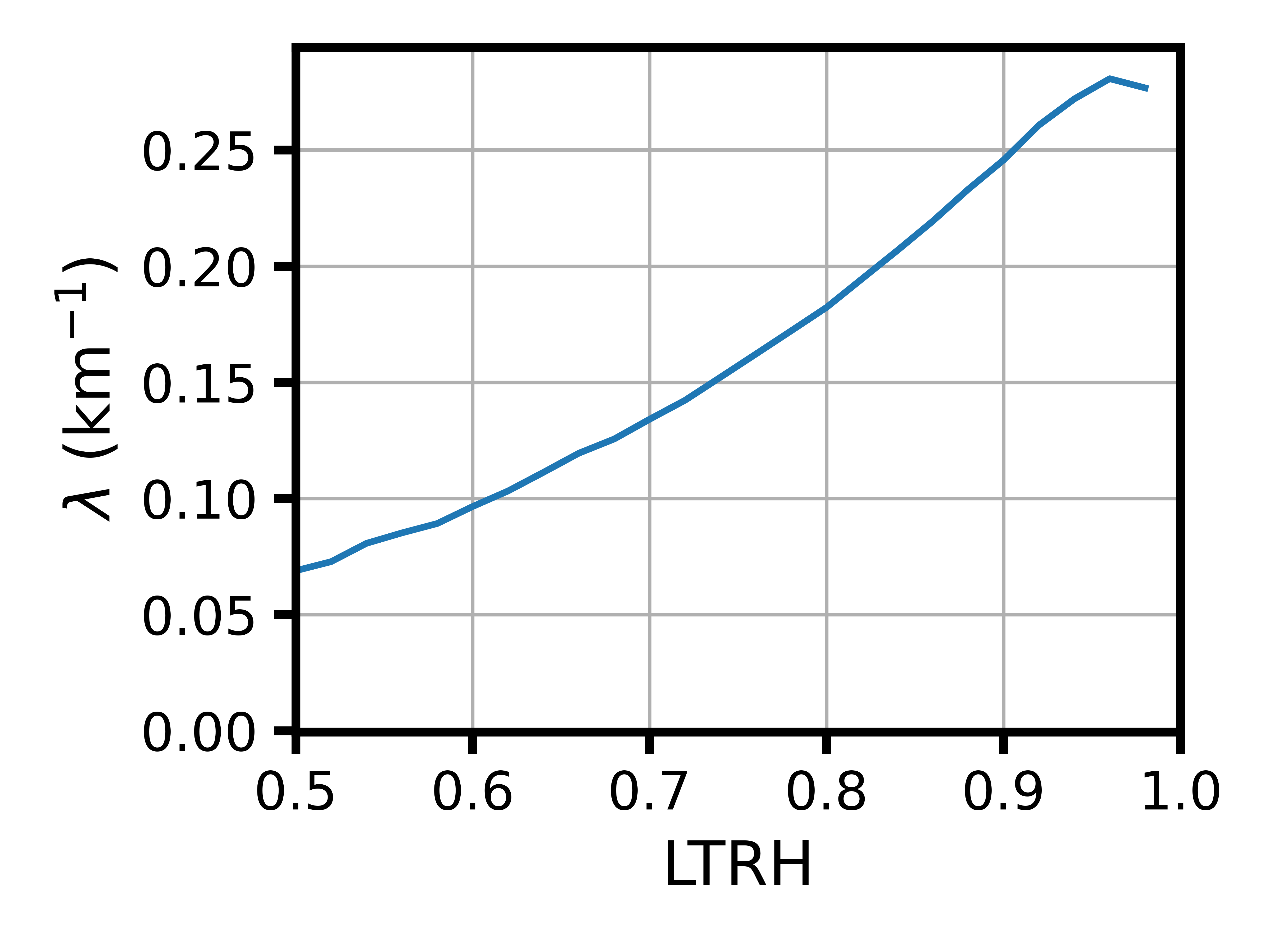}
    \caption{The dilution rate $\lambda$ from the stochastic parcel model, representing the average entrainment rate of non-detraining parcels between 1000 m and 4000.}
    \label{fig:dilution}
\end{figure}

\begin{linenomath*}
\begin{equation}
\lambda(z) = -\frac{log\ \bar{\phi_p}}{z - z_b}
\end{equation}
\end{linenomath*}

where $\bar{\phi_p}$ is the mass-flux-weighted average of $\phi_p$ and $z_b = 1\ km$ is the presumed altitude of the cloud base. One should think of the dilution rate $\lambda$ as the average entrainment rate of non-detraining parcels.

Figure \ref{fig:dilution} shows how $\lambda$ evaluated at 4 km depends upon $LTRH$. Convectively active air become more diluted as environmental humidity increases. This result suggests that the bulk entrainment rate appropriate for a plume model depends on the humidity. This is not an entirely novel result. \citeA{Becker2021} found that the rate of bulk plume entrainment (i.e., dilution) increased with humidity in a cloud-resolving simulation of the tropics. Those authors attributed their result to buoyancy sorting, as heavily entraining parcels of air are preferentially detrained in dry environments. The parcel model here suggests the same interpretation. Other studies have shown that, assuming convective quasi-equilibrium, greater lower-free troposphere humidity is associated with greater bulk plume entrainment rates \cite{Ahmed2020,Emmenegger2024}. However, the parcel model makes no explicit equilibrium assumption, allowing for us to draw a causal inference: Greater humidity causes greater dilution.

\subsection{The buoyancy-humidity relationship in the parcel model}

Buoyancy ($T_v'$) exhibits two behaviors in the GSRMs that we wish to better understand: a muted sensitivity of 600 hPa $T_v'$ to $LTRH$ and a modest decline in $T_v'$ beyond the modal value of $LTRH$. In order to understand the dependence of mid-level $T_v'$ on $LTRH$, we again turn to the parcel model. 

\begin{figure}
    \centering
    \includegraphics[width=0.5\linewidth]{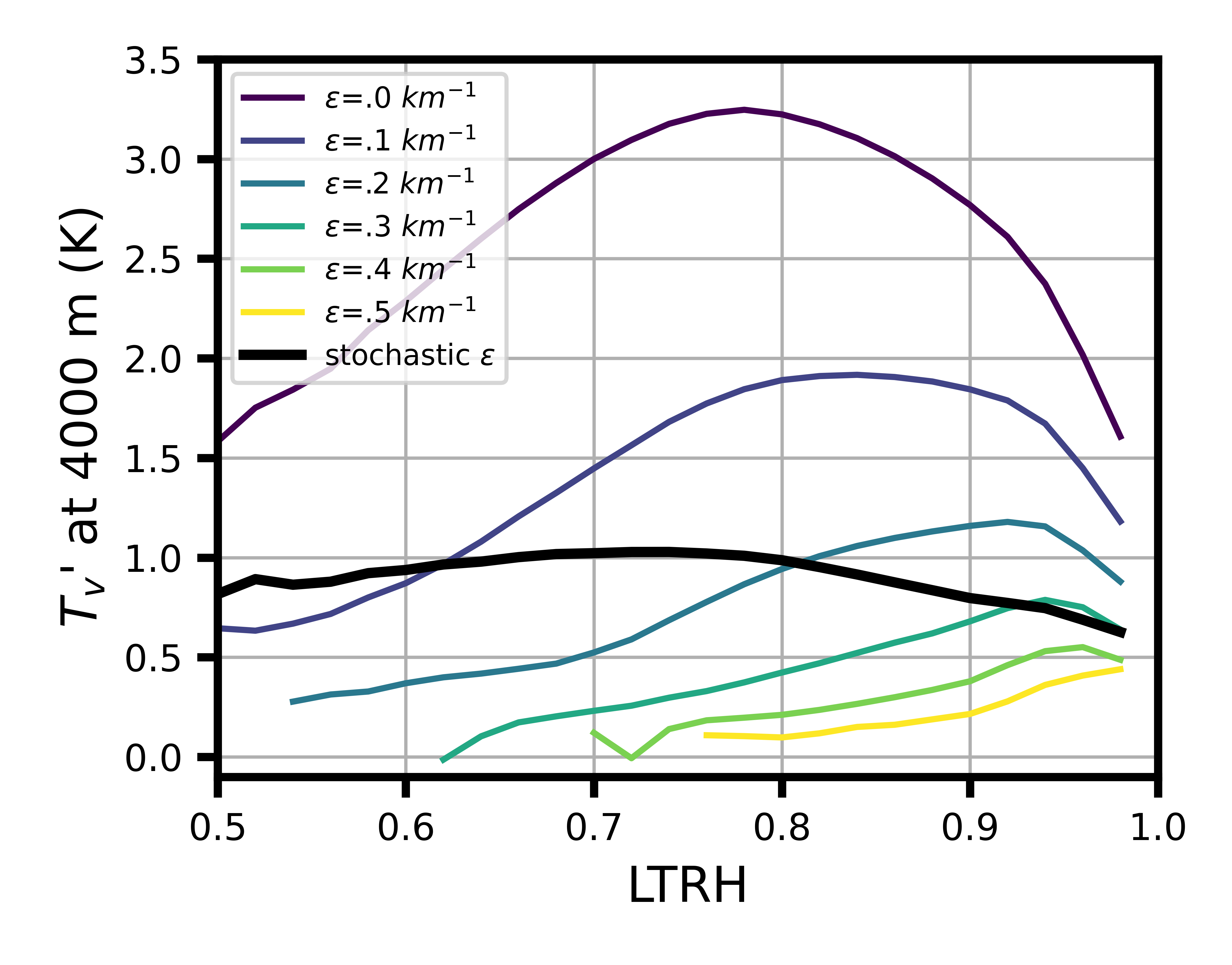}
    \caption{Buoyancy, $T_v'$, as measured by the difference between the virtual temperature of convectively active air and the virtual temperature of the environment at 4000 m in the parcel model.}
    \label{fig:tvprime_parcels}
\end{figure}

Fig. \ref{fig:tvprime_parcels} displays $4000\ m$ $T_v'$ from the parcel model. As with the GSRMs, the $LTRH$-bin averages employ mass flux weighting. With zero entrainment (darkest blue curve), $T_v'$ increases up to the modal $LTRH$ value in MERRA-2 ($LTRH \approx 0.8$, see Fig. \ref{fig:p_pickup}a), which approximates the convective critical point. Beyond the critical point, $T_v'$ declines. For surviving parcels, the nonentraining buoyancy has two components: the virtual temperature of the environment and the saturation virtual temperature of a moist adiabat, the latter of which is a function of boundary-layer moist static energy. Virtual temperature varies little with respect to humidity in the middle troposphere, as perturbations in humidity and temperature tend to cancel one another in order to maintain a weak virtual temperature gradient \cite{Yang2020,Bao2021,Seidel2024}. Therefore, the decline in non-entraining $T_v'$ after the critical point is likely due to a decline in the boundary-layer moist static energy as deep convection becomes more widespread. However, $T_v'$ in the non-entraining parcel model shows much larger sensitivity to $LTRH$ than was simulated by in the GSRMs (Fig. \ref{fig:tvprime_gsrms}). 

As we progress from lesser to greater fixed entrainment rates in the parcel model (Fig. \ref{fig:tvprime_parcels}, colored curves), $T_v'$ declines monotonically for any fixed $LTRH$. This is expected, for greater entrainment will lead to a greater reduction in $T_v'$ at any fixed relative humidity. At fixed entrainment rates of $\epsilon = 0.2\ km^{-1}$ or greater, which might be appropriate for tropical deep convection through the lower troposphere, the value of $T_v'$ generally increases with $LTRH$ past the critical point. This behavior is opposite from what we saw in the GSRMs, in which $Tv'$ declines beyond the critical point. Only stochastic entrainment (Fig. \ref{fig:tvprime_parcels}, black curve), produces both the muted sensitivity of $T_v'$ to humidity before the critical point and a modest decline in $T_v'$ after it. 

 What explains the muted response of population-mean $T_v'$ to humidity in the stochastic parcel model? This is a consequence of dilution. The greater dilution rate in humid environments reduces the mean $T_v'$ of convectively active air, dampening the positive trend in $T_v'$ with humidity which we saw for large, fixed entrainment rates. Another way to understand the muted response of $T_v'$ is as a direct consequence of buoyancy sorting and heterogeneous entrainment. We can think of buoyancy sorting as a form of natural selection. To survive, a parcel must maintain non-negative buoyancy, and the average properties of survivors will always reflect that zero-buoyancy threshold. Furthermore, parcels with near-zero buoyancy tend to dominate the population of survivors. Because the modal entrainment rate is so large (Fig. \ref{fig:distributions}c, black curve), an outsize share of surviving mass flux is composed of marginal survivors: those with \textit{just enough} buoyancy to survive. This drives the average $T_v'$ of ascending air towards zero and offsets the greater $T_v'$ of non-marginal parcels. The humidity-dependence of the dilution rate (shown in Fig. \ref{fig:dilution}) is largely a consequence of this drive towards zero buoyancy.

So far we have offered an explanation only for why it makes sense for there to be \textit{little} change in convective buoyancy with increasing humidity. While we understand why the buoyancy-sorting process causes the dilution rate to increase with $LTRH$, it is not obvious why the dilution rate increases so much that there is a strictly \textit{negative} trend in buoyancy beyond the mode of the $LTRH$ distribution. It is encouraging that a simple parcel model can capture this behavior, but a satisfying explanation has eluded us thus far. Therefore, we can only offer a speculative explanation in the following paragraphs. 

There is a one-to-one relationship between the virtual temperature of a saturated parcel and its moist static energy, which is conserved during non-entraining moist-adiabatic ascent. Due to buoyancy sorting, the moist static energies of parcels at any given level have been filtered according to the virtual temperature of the environment below. If environmental virtual temperatures below are relatively large, then the requirement for survival is that parcels have a relatively large moist static energy. Thus the mean virtual temperature of a population of stochastic \textit{parcels} should be positively correlated with the mean virtual temperature of the \textit{environment} below. Buoyancy, determined by the difference in virtual temperature between a parcel and its immediate environment, should then correlate with the vertical gradient in virtual temperature. A greater lapse rate in environmental virtual temperature should correlate with greater buoyancy. Accordingly we calculate the bulk virtual temperature lapse rate between 2 km and 4 km altitude in the MERRA-2 data:

\begin{linenomath*}
\begin{equation}
\Gamma_v = -\frac{T_v(z = 4 km) - T_v(z = 2 km)}{2 km}
\end{equation}
\end{linenomath*}

\begin{figure}
    \centering
    \includegraphics[width=0.5\linewidth]{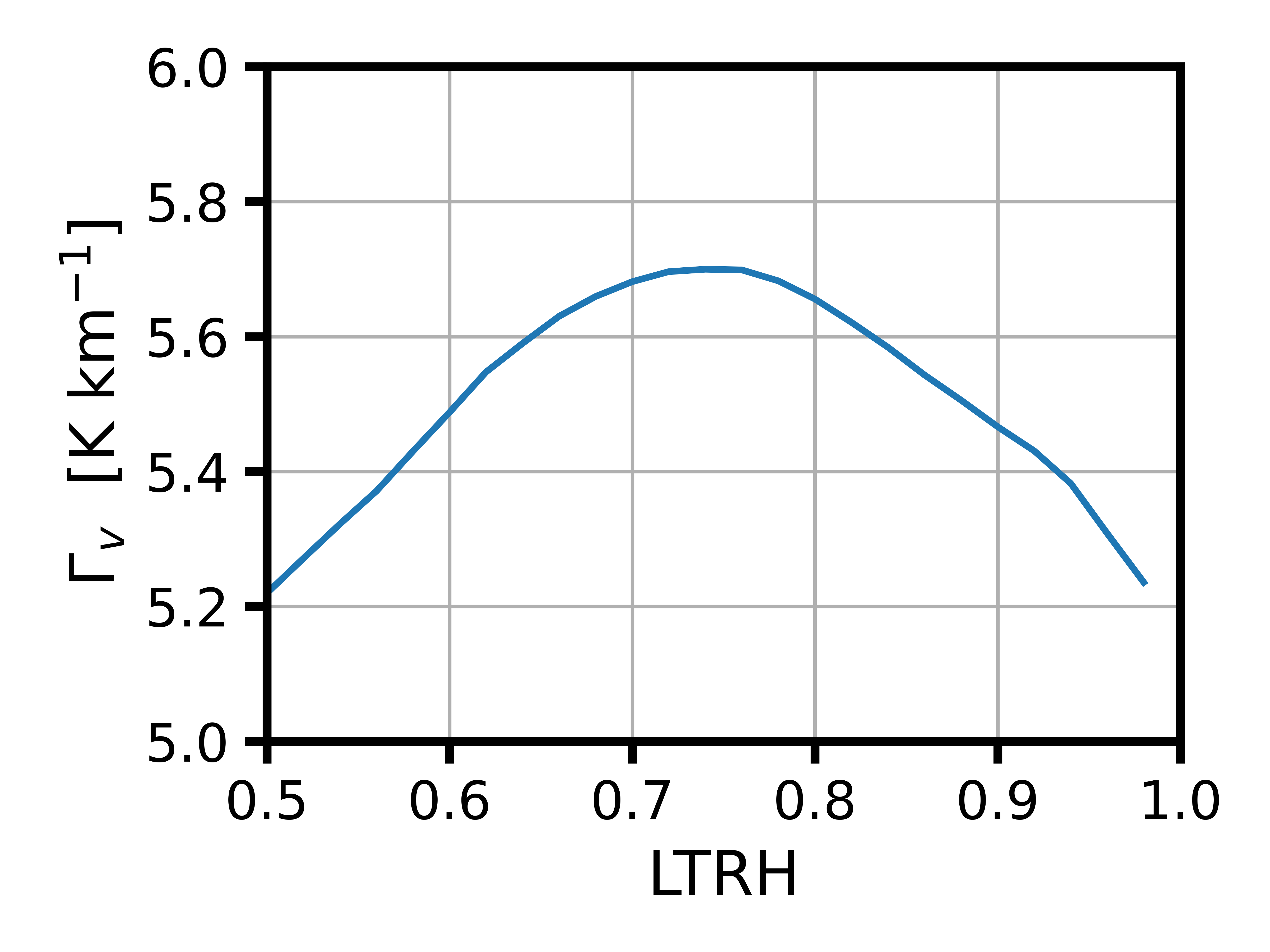}
    \caption{The bulk virtual temperature lapse rate, $\Gamma_v$, between 2 km and 4 km altitude in the MERRA-2 data.}
    \label{fig:lapse}
\end{figure}

Figure \ref{fig:lapse} shows that $\Gamma_v$ rises and then declines with $LTRH$, similar to $Tv'$ the stochastic parcel model. Based on the similarity between MERRA-2's $\Gamma_v$-$LTRH$ relationship and the stochastic parcel model's $T_v'$-$LTRH$ relationship, the speculative explanation we offered above may be correct: The moist static energies of parcels, and thus their present virtual temperatures, are filtered so that the averages of these quantities correlate with the environmental virtual temperature below the parcels' present level. Using any lower bound between 0 and 2 km when calculating $\Gamma_v$ results in a qualitatively similar $\Gamma_v$-$LTRH$ relationship. We have chosen the 2 km level because the relevant vertical scale over which $T_v'$ is determined is also likely to be small, due to the relatively short return distance of stochastic entrainment ($\alpha = 500\ m$). Our reader should bear in mind that we have not given a quantitative theory for translating a trend in $\Gamma_v$, or any other variable, into a trend in $T_v'$. Furthermore, the explanation we offer here has proven difficult to test in the parcel model, as we would need to specify what measure of humidity should remain fixed in order to isolate the effect of varying the environmental lapse rate. This remains an open problem.

\subsection{Two-layer analysis of parcel-model mass flux}

\begin{figure}
    \centering
    \includegraphics[width=1\linewidth]{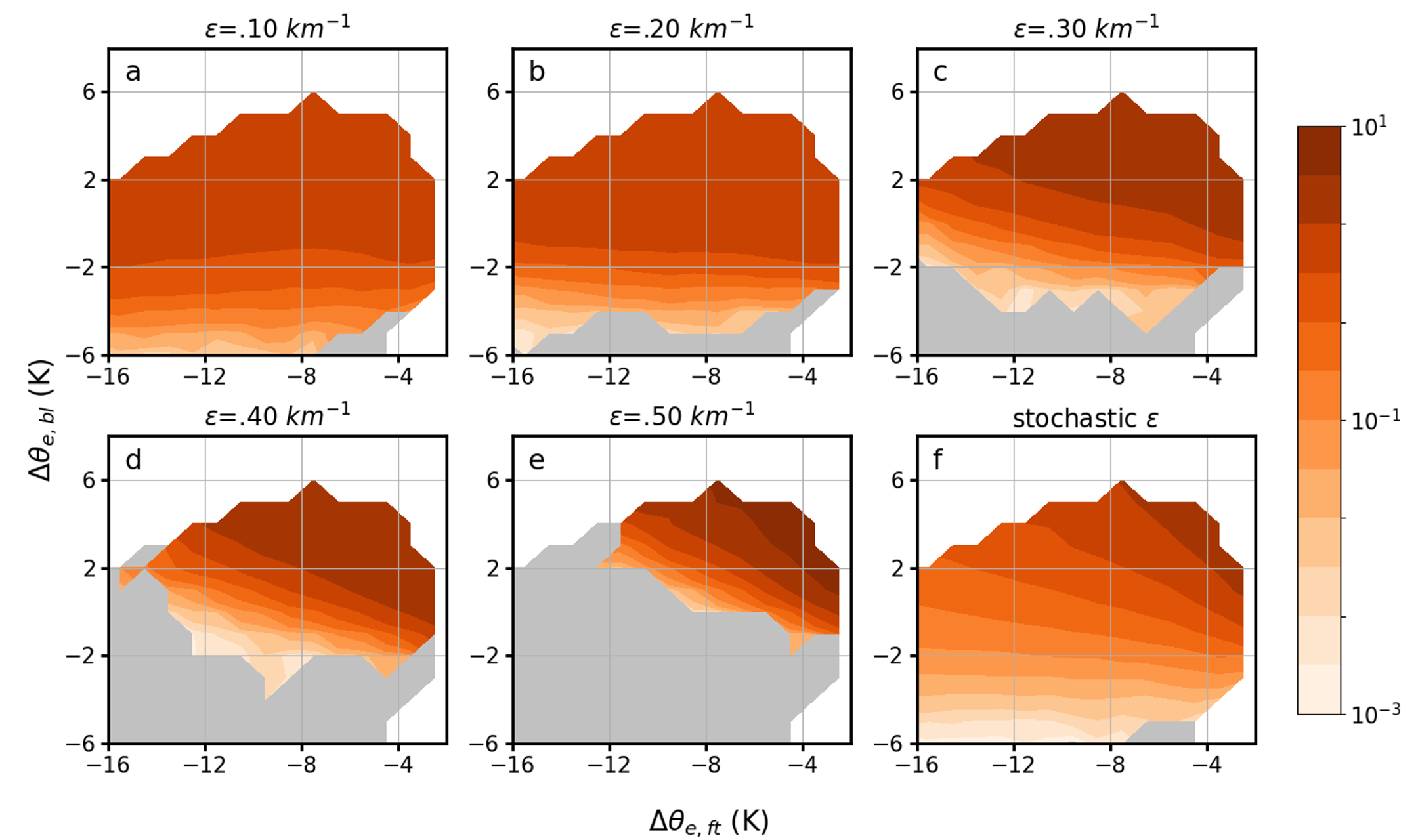}
    \caption{Parcel-model mass fluxes at 4000 m conditionally averaged on $\Delta \theta_{e,ft}$ and $\Delta \theta_{e,bl}$ for various entrainment rates: (a) $\epsilon = 0.1\ km^{-1}$, (b) $\epsilon = 0.2\ km^{-1}$, (c) $\epsilon = 0.3\ km^{-1}$, (d) $\epsilon = 0.4\ km^{-1}$, (e) $\epsilon =\ 0.5\ km^{-1}$, (f) stochastic $\epsilon$. Grey indicates where no parcels survived to 4000 m.}
    \label{fig:m_dth_parcels}
\end{figure}

The stochastic parcels also replicate the qualitative behavior of the GSRMs when mid-level mass flux is binned by $\Delta \theta_{e,ft}$ and $\Delta \theta_{e,bl}$, as in Section \ref{sec:twolayer}. Fig. \ref{fig:m_dth_parcels} shows the parcel mass flux at 4000 m conditionally averaged on $\Delta \theta_{e,ft}$ and $\Delta \theta_{e,bl}$ for several fixed rates of entrainment as well as for stochastic entrainment. Each fixed rate of entrainment fails to fully replicate the behavior of the GSRMs in Fig. \ref{fig:m_dth}. At smaller rates of entrainment, the parcel model is insufficiently sensitive to increases in $\Delta \theta_{e,ft}$. At greater rates of entrainment, the parcel model is unable to simulate the convection present at low values of $\Delta \theta_{e,bl}$. 

Stochastic entrainment resolves both of the above-mentioned difficulties (Fig. \ref{fig:m_dth_parcels}f). The stochastic parcel model is able to qualitatively replicate the mass flux in the GSRMs (Fig. \ref{fig:m_dth}). The contours in Fig. \ref{fig:m_dth_parcels}f are horizontal near the bottom region of the plot but become more tilted in the upper right region. In physical terms, the mass flux is primarily sensitive to $\Delta \theta_{e,bl}$ in the least convective environments but becomes more sensitive to $\Delta \theta_{e,ft}$ in the most convective environments. This is due to the dependence of the dilution rate on the environment. In the most humid environments, which have greater $\Delta \theta_{e,ft}$ and $\Delta \theta_{e,bl}$, the dilution rate is greater. This introduces greater sensitivity to  $\Delta \theta_{e,ft}$, as more of the convectively active air originated in the free troposphere environment. The qualitative similarity to the GSRMs suggests that the stochastic parcel model accurately emulates the physics governing bulk mass flux in the GSRMs. 

\section{Increase in updraft area fraction with humidity}\label{sec:area}

\begin{figure}
    \centering
    \includegraphics[width=0.75\linewidth]{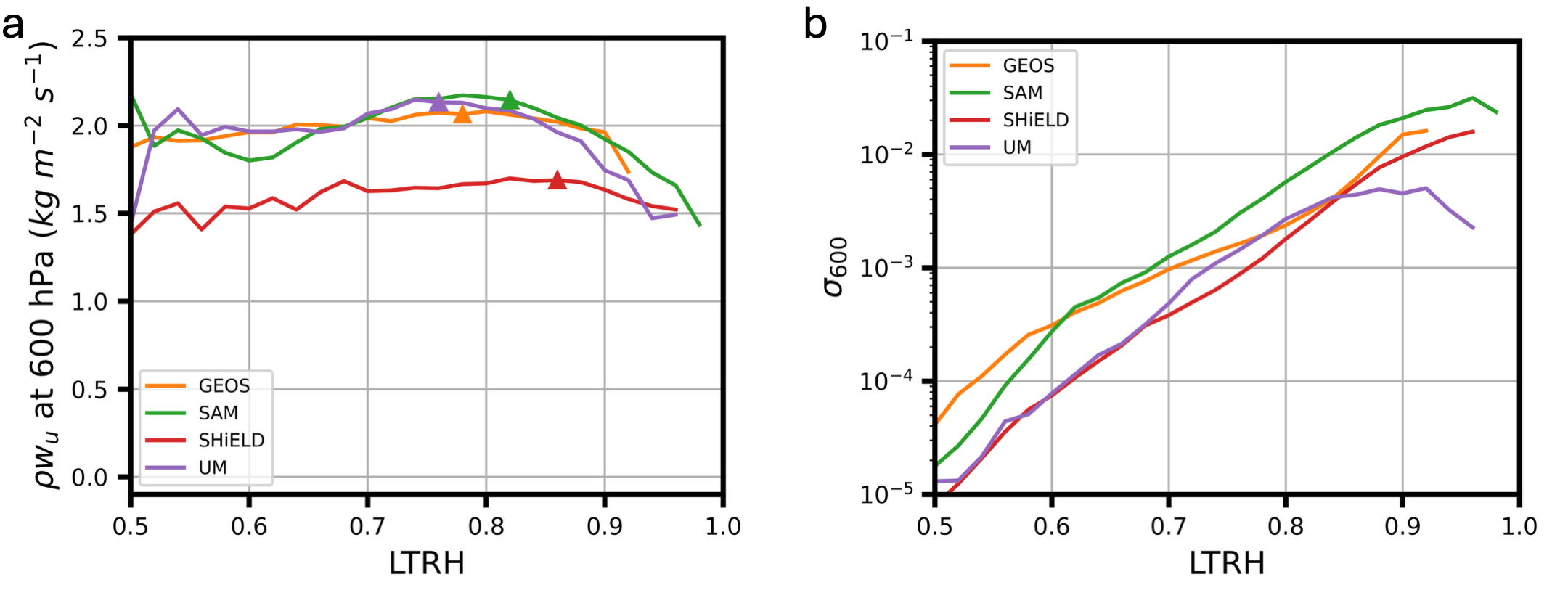}
    \caption{(a) Mean updraft velocity at 600 hPa in the GSRMs. (b) Updraft area fraction at 600 hPa in the GSRMs.}
    \label{fig:area_velocity}
\end{figure}

The pickup in mid-level mass flux with humidity must be due to a pickup in updraft velocity and/or a pickup in updraft area (See Eq. 1). We shall investigate which of these two is dominant. In the previous sections, we showed that updraft buoyancy declines with increasing humidity. We would expect that updraft velocity is correlated with buoyancy, as the upward velocity in convection is generally the result of buoyant accelerations. The decline in convective buoyancy with increasing $LTRH$ should cause updraft velocity to decline as well. Figure \ref{fig:area_velocity}a shows that the GSRMs corroborate this intuition: Updraft velocity declines above the maximum in the $LTRH$ distribution all five GSRMs we analyzed. 

If updraft velocity declines with increasing $LTRH$, then the pickup in mass flux must be due to a pickup in the area occupied by updrafts. Figure \ref{fig:area_velocity}b shows the relationship between $LTRH$ and the fractional area occupied by updrafts in the GSRMs. For each model, the pickup in updraft area is qualitatively indistinguishable from the pickup in mass flux in Fig. \ref{fig:m_pickup}b. Our reader should bear in mind that this is only a measure of \textit{total} updraft area. These data are silent regarding whether the pickup in mass flux is due to individually wider updrafts or a greater population of updrafts.

\begin{figure}
    \centering
    \includegraphics[width=0.5\linewidth]{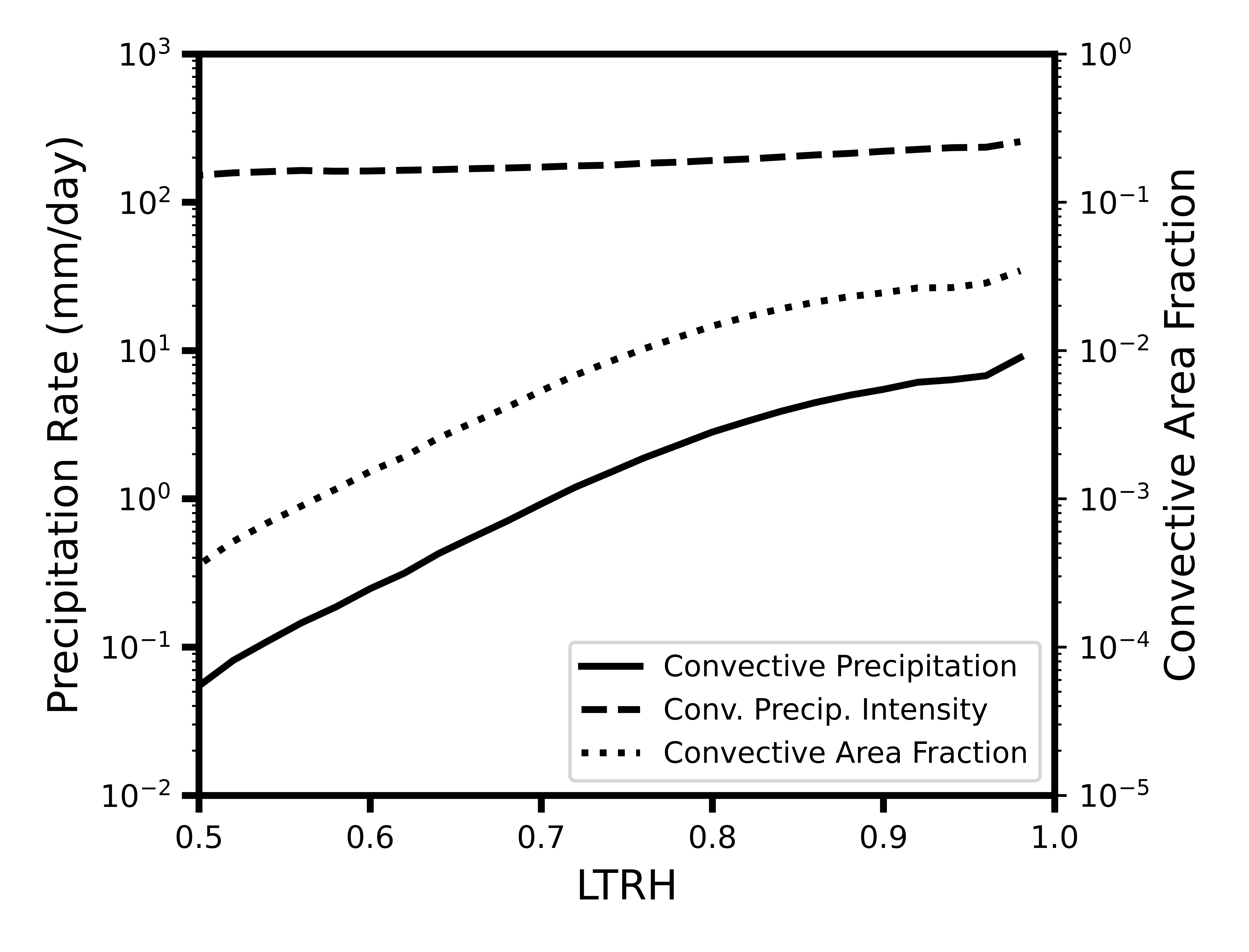}
    \caption{Convective precipitation, convective area fraction, and the intensity of convective precipitation in TRMM for the period 2010-2013. LTRH values are derived from MERRA-2 reanalysis.}
    \label{fig:trmm}
\end{figure}

We now present an observational test of this behavior. We use data from the Tropical Rainfall Measurement Mission (TRMM) which partition the total rainfall between convective, stratiform, and shallow modes of precipitation. These classifications are performed at the horizontal scale of a 5 km pixel \cite{Funk2013,Ahmed2015}, allowing us to provide both an approximate convective area fraction and a conditional rainfall intensity. Figure \ref{fig:trmm} shows total convective precipitation, convective area fraction, and the local intensity of convective precipitation as binned by MERRA-2-derived $LTRH$. The convective precipitation (solid line) has nearly the same trend as convective area fraction (dotted line). The observed pickup in convective precipitation is primarily due to a pickup in convective area fraction.

The intensity of convective rain when convection is present (dashed line in Fig. \ref{fig:trmm}) increases somewhat over the range of LTRH. If we take the intensity as a proxy for the vertical velocities associated with convection, then this result is similar to the GSRMs. The GSRMs show a slight decline in vertical velocity with $LTRH$, and TRMM shows a slight increase in conditional rainfall intensity. To first order, both results indicate a relatively small range in convective intensity relative to convective area. Furthermore, TRMM's positive trend in precipitation intensity may also be explained by a positive trend in updraft area within the 5 km-wide TRMM pixels rather than a pickup in intensity at the updraft scale. Ground-based radar profilers and satellite-based estimates have similarly shown that updraft area dominates the observed variability in convective mass flux \cite{Kumar2016,Giangrande2016,Jeyaratnam2021}. In the future, it would be interesting to link those studies and this one by characterizing the response of updraft area to humidity using observations.

\section{Conclusion: survival and dilution govern how convective mass flux responds to environmental humidity}\label{sec:conclusion}

We have documented the following three relationships between lower tropospheric relative humidity ($LTRH$) and convection in the DYAMOND GSRMs:

\begin{itemize}
  \item R1: Convective mass flux increases with height through the lower free troposphere in humid environments, but declines with height in dry environments.
  \item R2: Mid-level convective mass flux increases nonlinearly with $LTRH$, exhibiting a quasi-exponential relationship similar to the well-documented pickup in observed precipitation.
  \item R3: The mean buoyancy of convection exhibits a muted response to $LTRH$ and even declines beyond the mode of the $LTRH$ distribution.
\end{itemize}

We describe a simple parcel model, which we use to identify a minimum recipe for these three relationships. That recipe has two key ingredients, buoyancy sorting and stochastic entrainment. The combination of these two ingredients produces the survival and dilution effects. At greater humidity, a greater fraction of parcels survive to ascend rather than detrain, and those survivors are on average more diluted by environmental air. These explain the three relationships between humidity and convection in the GSRMs. The deep inflow mass flux profile (R1) forms when the rate of survival is great enough that entrainment exceeds detrainment through the depth of the lower free troposphere. Both survival and dilution cause mass flux to increase nonlinearly with humidity (R2). And a pickup in dilution is partly responsible for the \textit{decline} in convective buoyancy (R3), though there are likely other factors at work.

We believe that the decline in convective buoyancy at high humidity (R3) is especially consequential for our understanding of tropical convection. When designing numerical simulations or outlining theories involving convection, it is useful to relate the convective mass flux of a population of clouds (or their resulting precipitation) to the large-scale thermodynamic environment. Many models of convection accomplish this by linking the magnitude of precipitation or convective mass flux to the buoyancy of a representative entraining plume \cite{Arakawa1974,Ahmed2020,Wolding2024}. These models have generally succeeded. A properly tuned buoyancy estimate is a useful relative measure of how hospitable the large-scale environment is for deep convection. Environments with greater entraining plume buoyancy tend to have greater convection. However, the true buoyancy of convective updrafts - as estimated in this study - behaves in the opposite manner: It declines with humidity.

Given that the buoyancy in a plume-buoyancy model is not correct, how does it succeed in identifying convective environments? This is due to the relationship between plume buoyancy and the survival effect. Parcels are sorted out of a cloud based on their buoyancy, which naturally depends upon the same environmental variables as the buoyancy of a simple plume with fixed entrainment. However, an idealized plume buoyancy can only predict which environments are relatively more or relatively less supportive of convection. Without an additional layer of parametrization which accounts for the effects of survival and dilution, plume buoyancy cannot provide a quantitative estimate of convective mass flux. In convection schemes, this gap is often closed by prescribing cloud-base mass flux to scale linearly with a vertical integral of buoyancy \cite{Arakawa1974,Betts1986,Zhang1995}. In contrast, the stochastic parcel model does not prescribe the cloud-base mass flux to respond to the environment aloft. By way of the survival and dilution processes, the stochastic parcel model allows the environment to directly influence the convective mass flux of cloudy air throughout its ascent. The stochastic parcel model is a more physically plausible model of convection. It does not require parcels at the cloud base to ``know" about the environmental conditions aloft.

The success of the stochastic parcel model suggests that convection parameterizations would be more physically realistic if they employ varying entrainment rates. One obvious choice to achieve this would be to implement a stochastic parcel parameterization such as the one described by \citeA{Romps2016}. Other schemes rely upon multiple plumes to simulate a diversity of entrainment rates \cite{Arakawa1974,Peters2021}. If each plume in a scheme has a fixed entrainment rate, it is likely that the correct linear combination of plumes could create both a quasi-exponential pickup in mass flux with humidity and a pickup in the dilution rate (i.e., bulk entrainment rate). We hope that our results in this study are a useful guide when evaluating these and other convection schemes.

\section{Discussion: limitations and opportunities}\label{sec:discussion}

For this study we have calculated convective mass flux fields from the output of five global storm-resolving models (GSRMs). On the one hand, their mass fluxes tell a consistent story, which is the focus of this paper. On the other hand, these mass fluxes are quantitatively very different from model to model. The same is true for buoyancy, $T_v'$. When we were preparing the parcel model, our tuning tests taught us that even modest changes to entrainment assumptions or to environmental conditions can cause substantial quantitative differences in mass flux. This may present an opportunity to better understand the GSRMs: can a parcel model capture those intermodel differences? Another avenue to understand the intermodel spread would be to quantify their static stabilities and radiative cooling rates. Together, these two variables regulate the total mass flux, and static stability in particular may vary considerably between models \cite{Williams2025,Stauffer2022}. 

Another significant opportunity for future work derives from the parcel model's prediction that deep convective mass flux is more diluted where environmental humidity is greater. If true, this fact would inform our understanding not only of the pickup of convection with humidity, but on the proper design of convection parameterizations. We are currently testing this prediction in large-eddy simulation experiments. Such simulations may also help to test whether the mass flux pickup with environmental humidity is the result of heterogeneous entrainment, as we have suggested in this paper, or if it is rather the result of  sub-grid-scale heterogeneity in the thermodynamic fields.

One shortcoming of this study is the coarse resolution at which convection is simulated in the GSRMs. Although a resolution of 2-5 km is small enough that a deep convective parameterization is not needed, that resolution is of the same order as the width of a deep convective updraft (1-3 km). Consequently, the simulated updrafts may be unrealistically wide, with a rate of entrainment which is too small \cite{Varble2014}. The in-cloud dynamics which may influence entrainment and detrainment may also be partly absent. We hope that our conclusions are made more convincing by the agreement between the parcel model and the GSRMs and between the GSRMs and observations. However, one may worry that the parcel model is more useful for emulating km-scale convection-permitting simulations than actual convection. Here, too, is an opportunity for follow-up work with convection-\textit{resolving} large-eddy simulations. We also believe this is an opportunity for the upcoming INCUS mission to measure deep convective updraft velocity from orbit (https://incus.colostate.edu/). Will observationally derived convective mass flux tell the same story as the models?

\section*{Open Research Section}
The DYAMOND GSRM data are stored on DKRZ's Levante HPC system, publicly available to registered users \cite{Stevens2019}.  The MERRA and ERA reanalysis and TRMM and IMERG precipitation products are publicly available from their respective providers \cite{GlobalModelingandAssimilationOffice2015,Hersbach2020,Funk2013,Huffman2023}. Processed GSRM data and observations are available on Zenodo: https://doi.org/10.5281/zenodo.15694788 \cite{Seidel2025}.

\acknowledgments
SS was supported by a NASA Postdoctoral Program fellowship. NA was supported by the NASA Modeling, Analysis, and Prediction Program. BW was supported in part by the NSF (AGS-2225957) and in part by NOAA Cooperative Agreement NA22OAR4320151. The manuscript benefited from thoughtful reviews by John Peters and two anonymous reviewers, and it benefited from helpful editorial suggestions by Kathleen Schiro. The authors thank Fiaz Ahmed for providing the regridded TRMM data used in this study. This research would not have been possible without computational resources provided by the NASA Center for Climate Simulation (NCCS). DYAMOND data management was provided by the German Climate Computing Center (DKRZ) and supported through the projects ESiWACE and ESiWACE2. The projects ESiWACE and ESiWACE2 have received funding from the European Union’s Horizon 2020 research and innovation programme under grant agreements No 675191 and 823988. This work used resources of the Deutsches Klimarechenzentrum (DKRZ) granted by its Scientific Steering Committee (WLA) under project IDs bk1040 and bb1153. 

\section*{Conflicts of Interest}

The authors declare there are no conflicts of interest for this manuscript.

\bibliography{MassFluxHumidity}

%
%


%
%
%
%
%

\end{document}